\documentclass[sigconf, nonacm]{acmart}

\newcommand\vldbdoi{XX.XX/XXX.XX}
\newcommand\vldbpages{XXX-XXX}
\newcommand\vldbvolume{14}
\newcommand\vldbissue{1}
\newcommand\vldbyear{2020}
\newcommand\vldbauthors{\authors}
\newcommand\vldbtitle{\shorttitle}
\newcommand\vldbavailabilityurl{URL_TO_YOUR_ARTIFACTS}
\newcommand\vldbpagestyle{plain}

\newtheoremstyle{bolddef} 
  {}                      
  {}                      
  {}                      
  {}                      
  {\bfseries}             
  {.}                     
  { }                     
  {\thmname{#1}\thmnumber{ #2}\thmnote{ (#3)}} 

\theoremstyle{bolddef}

\newtheoremstyle{italicex}
  {} 
  {} 
  {\itshape} 
  {} 
  {\bfseries} 
  {.} 
  {.5em} 
  {} 

\theoremstyle{italicex}
\newtheorem{definition}{Definition}
\newtheorem{example}{Example}

\newtheorem{theorem}{Theorem}
\newtheorem{lemma}{Lemma}

\usepackage{graphicx}
\usepackage{booktabs}
\usepackage[skip=5pt]{caption}
\usepackage{subcaption}
\usepackage{multirow}
\usepackage{weiwAlgorithm}
\usepackage{amsmath}
\usepackage{newtxmath}
\usepackage{diagbox}
\usepackage{url}
\usepackage{dsfont}
\usepackage{enumitem}
\usepackage{caption}
\usepackage{balance}
\usepackage{tabularx}
\usepackage{svg}
\usepackage{xcolor}
\usepackage{float}
\usepackage[normalem]{ulem}
\usepackage{appendix}
\usepackage{colortbl}
\definecolor{fullrow}{HTML}{DDE6F7}

\newcommand{\myparagraph}[1]{\vspace{1mm} \noindent \textbf{#1}.}

\newcommand{\supplementref}[2]{%
  \ifcsname r@#1\endcsname\ref{#1}\else#2\fi}

\begin{document}

\title{KaleidoKV: Rearranging Evidence for Efficient KV Cache Reuse in Offline RAG Inference}

\author{Peiting Xie}
\affiliation{%
  \institution{The University of New South Wales}
  \city{Sydney}
  \state{Australia}
  \postcode{2032}
}
\email{peiting.xie@unsw.edu.au}

\author{Xingyu Tan}
\affiliation{%
  \institution{The University of New South Wales}
  \city{Sydney}
  \country{Australia}
}
\email{xingyu.tan@unsw.edu.au}

\author{Xiangjun Zai}
\affiliation{%
  \institution{The University of New South Wales}
  \city{Sydney}
  \country{Australia}
}
\email{xiangjun.zai@unsw.edu.au}

\author{Chen Chen}
\affiliation{%
  \institution{University of Shenzhen}
  \city{Shenzhen}
  \country{China}
}
\email{chencc@szu.edu.cn}

\author{Xiaoyang Wang}
\affiliation{%
  \institution{The University of New South Wales}
  \city{Sydney}
  \country{Australia}
}
\email{xiaoyang.wang1@unsw.edu.au}

\author{Wenjie Zhang}
\affiliation{%
  \institution{The University of New South Wales}
  \city{Sydney}
  \country{Australia}
}
\email{wenjie.zhang@unsw.edu.au}

\newcommand{\sysname}{KaleidoKV}

\begin{abstract}
\begingroup
\setlength{\parfillskip}{0pt plus 0.08\linewidth}
\tolerance=1000
\emergencystretch=0.5em
\finalhyphendemerits=10000
Retrieval-augmented generation (RAG) augments large language models (LLMs) on question-answering (QA) tasks with retrieved evidence at inference time. 
Related questions in an offline batch often retrieve overlapping evidence, but prefix caching reuses key-value (KV) states only for identical token prefixes, making reuse dependent on evidence order. Although all questions and evidence are known before execution, existing reordering methods commit to early grouping decisions and can miss sharing opportunities that require several requests to move together. Evaluating alternative layouts can incur planning costs that offset inference savings, while concurrent requests may still recompute shared prefixes before their KV states become available.
To address these challenges, we propose \textbf{KaleidoKV}, a framework that plans evidence orders across the batch and controls when requests enter the engine while preserving each request's evidence set and content. 
Joint layout planning revises formed groups through coordinated request moves and keeps a move only if it lowers planned prefill, the computation required when each shared prefix is computed once. A global prefix index selects evidence orders by locating reusable prefixes and evaluates each move incrementally from only the affected request paths. Prefix-ready admission tracks nested prefix dependencies, delaying requests whose shared prefixes are still being computed while allowing independent requests to proceed.
Experiments on four QA workloads and four LLMs show that KaleidoKV achieves the highest throughput and lowest mean time to first token in all 16 settings, with approximately 1.24-1.75$\times$ the throughput of the strongest baseline, including planning time. With Llama-3.1-8B under the default configuration, actual prefill stays within 0.5\% of planned prefill across all four workloads.
\par
\endgroup
\end{abstract}

\maketitle
\begin{figure}[!t]
    \centering
    \includegraphics[width=\columnwidth]{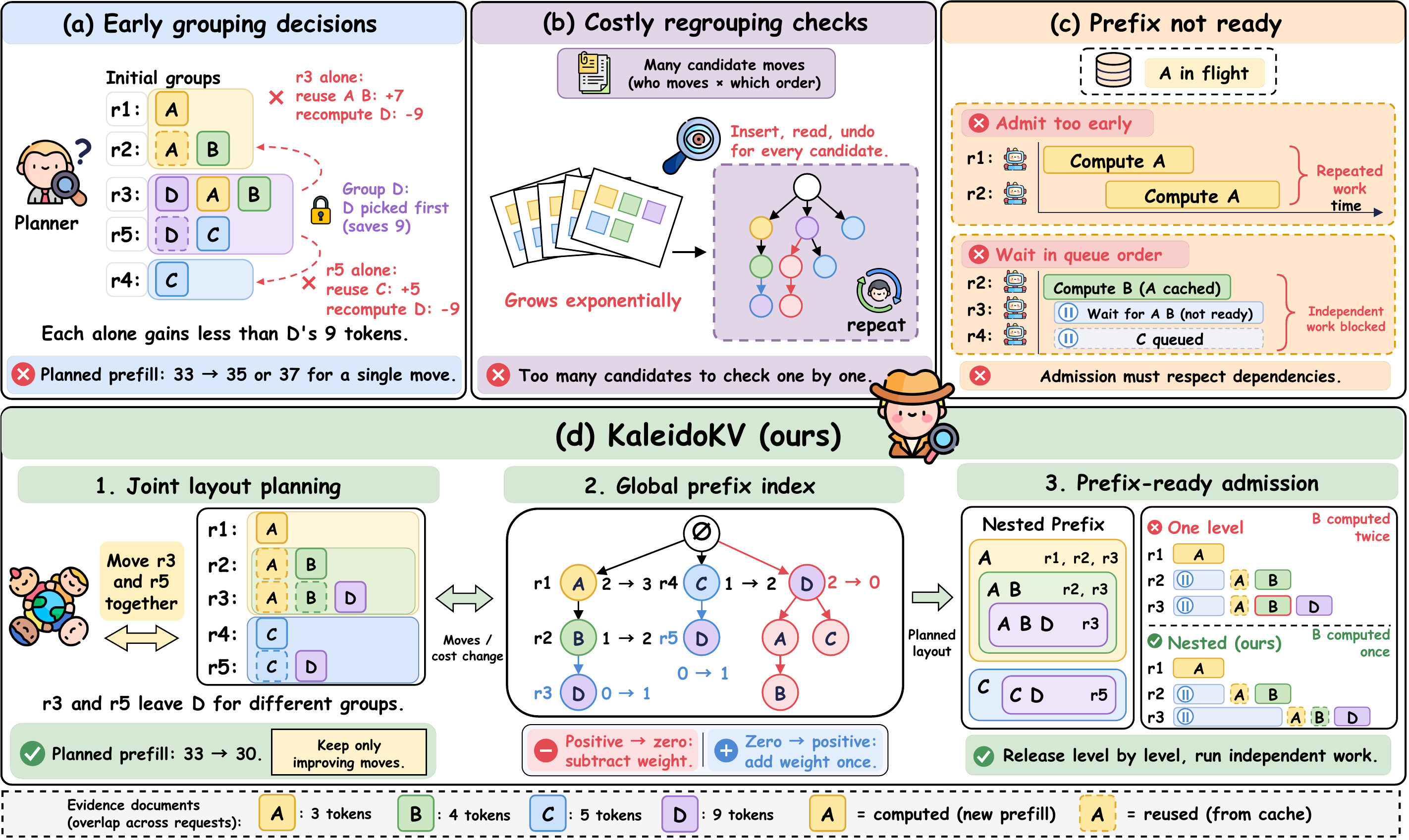}
    \caption{Challenges in existing work and KaleidoKV}
    \label{fig:intro}
\end{figure}

\pagestyle{\vldbpagestyle}
\vspace{-2mm}
\begingroup\small\noindent\raggedright\textbf{PVLDB Reference Format:}\\
\vldbauthors. \vldbtitle. PVLDB, \vldbvolume(\vldbissue): \vldbpages, \vldbyear.\\
\href{https://doi.org/\vldbdoi}{doi:\vldbdoi}
\vspace{-2mm}

\endgroup
\begingroup
\renewcommand\thefootnote{}\footnote{\noindent
This work is licensed under the Creative Commons BY-NC-ND 4.0 International License. Visit \url{https://creativecommons.org/licenses/by-nc-nd/4.0/} to view a copy of this license. For any use beyond those covered by this license, obtain permission by emailing \href{mailto:info@vldb.org}{info@vldb.org}. Copyright is held by the owner/author(s). Publication rights licensed to the VLDB Endowment. \\
\raggedright Proceedings of the VLDB Endowment, Vol. \vldbvolume, No. \vldbissue\ %
ISSN 2150-8097. \\
\href{https://doi.org/\vldbdoi}{doi:\vldbdoi} \\
}\addtocounter{footnote}{-1}\endgroup

\ifdefempty{\vldbavailabilityurl}{}{
\begingroup\small\noindent\raggedright\textbf{PVLDB Artifact Availability: \\}
The source code, data, and/or other artifacts have been made available at\\
\url{https://github.com/PeitingXie/KaleidoKV}.
\endgroup
}
\vspace{-2mm}

\section{Introduction}

\begingroup
\setlength{\parfillskip}{0pt plus 0.08\linewidth}
\tolerance=1000
\emergencystretch=0.5em
\finalhyphendemerits=10000
Retrieval-augmented generation (RAG) augments large language models
(LLMs) with relevant information retrieved at inference time \cite{zhou2025depth, lewis2020retrieval, gao2023retrieval, fan2024survey, tan2025hydrarag, gutierrez2024hipporag, RAGPerf, li2025neutronrag,zhao2026retrieval}. In batch
question-answering (QA) over document collections, related questions often
retrieve the same evidence. The model processes this evidence during
prefill before generating each answer, so executing requests
independently repeats computation on shared passages. Prefix caching
reuses previously computed key--value (KV)
states~\cite{kwon2023vllm, zheng2024sglang, zhang2026polarkv, li2025hotprefix, wang2026buoy}, but requires the same
tokens in the same order from the start of the prompt. Identical
passages following different prefixes cannot directly share their
cached states \cite{wang2026prefix}.

Evidence reordering can turn overlapping passages into shared prefixes
across requests. Following \cite{liu2025optimizing, zheng2026batchllm, zhao2026blendserve}, we study offline batches in which all questions and
their retrieved evidence are known before execution, allowing evidence
orders to be chosen jointly. Existing reordering methods place common
evidence first in the same order and schedule group members close
together~\cite{jiang2026contextpilot, liu2025optimizing}.
The evidence orders form a layout, represented as a prefix tree.
Its planned prefill is the evidence computation
required when each shared prefix is computed once. Our objective is to reduce total planning and inference time while
\mbox{preserving} each request's original evidence set and content.

\begin{example}
\label{ex:intro}
\setlength{\parfillskip}{0pt plus 0.08\linewidth}
Consider the five requests in Figure~\ref{fig:intro}, where $A$, $B$,
$C$, and $D$ denote evidence items of 3, 4, 5, and 9 tokens. Greedy
grouping first places $D$ at the start of $r_3$ and $r_5$, since sharing
$D$ saves the most tokens, which gives a planned prefill of 33 in
Figure~\ref{fig:intro}(a). Moving $r_3$ to $(A,B,D)$ and $r_5$ to
$(C,D)$ in Figure~\ref{fig:intro}(d) instead lets them share with $r_2$ and $r_4$, which lower
planned prefill to 30.
\vspace{-2.2mm}
\vspace{-2.2pt}

\end{example}

\vspace{1mm}
\noindent\textbf{Challenges in existing work.}
Example~\ref{ex:intro} shows that grouping requests by the evidence that saves the most does not necessarily minimize prefill across the batch.
A request assigned to one shared prefix may
lose the reuse it could obtain in another group. Revising these
assignments requires comparing alternative layouts, and the time
spent comparing them can offset the inference savings. Concurrent
requests may still compute the same prefix before its cached states
are available.

\vspace{0.1mm}

\noindent\uline{Challenge 1: Inflexible grouping without joint refinement.}
Early grouping decisions can prevent requests from joining groups with more reuse. Clustering and greedy reordering commit requests to shared prefixes as they form groups ~\cite{jiang2026contextpilot, liu2025optimizing}, but a better layout may require several requests to leave their groups together. In Example~\ref{ex:intro}, moving $r_3$ or $r_5$ alone computes $D$ a second time,
which raises the planned prefill to 35 or 37, whereas moving both
empties the shared $D$ and lowers it to 30. Likewise, requests from different groups may gain only by moving onto a new shared prefix together, since a request that moves there alone pays for the whole prefix. A search that accepts only improving single-request moves cannot reach such layouts. Revisiting formed groups requires evaluating request moves jointly, including moves that are individually unfavorable.

\vspace{0.1mm}

\noindent\uline{Challenge 2: Costly evaluation of grouping changes.}
Planning cost becomes a further obstacle when formed groups must be
revised. Existing clustering and greedy planners compare request pairs
or rescan remaining requests to form
groups~\cite{jiang2026contextpilot, liu2025optimizing}.
Revisiting these groups adds joint moves whose gains depend on which
other requests use the same prefixes. A request leaving a shared prefix
removes its prefill cost only if no other request still uses it, and
several requests moving to a previously unused prefix add its cost only
once. The prefix trees of inference engines hold only the current
prompts~\cite{zheng2024sglang}, so evaluating a proposed move requires
inserting the new paths, reading the change, and then undoing it
(Figure~\ref{fig:intro}(b)). A
moved request with $k$ evidence items may also take any of $k!$ orders,
and its best order depends on where the other moved requests go.
Planning must compute the cost change of a move, and find the best order of each moved request
without trying all $k!$ orders.

\vspace{0.1mm}

\noindent\uline{Challenge 3: Redundant prefill under concurrent execution.}
Scheduling reordered requests close together does not ensure that their
shared prefixes are ready when concurrent execution begins.
Existing methods either submit requests without checking shared-prefix readiness ~\cite{jiang2026contextpilot, liu2025optimizing, li2026solo} or compute shared prefixes first but keep only one level of sharing \cite{zheng2026batchllm}.
With chunked prefill~\cite{agrawal2024taming}, $r_2$ in
Figure~\ref{fig:intro}(c) can enter the engine while $r_1$ is still
computing $A$, repeating that prefill. Once $A$ is cached, $r_3$ may
still need to wait while $r_2$ computes $B$. Holding every later request
behind $r_3$ also delays $r_4$, although its prefix $C$ is independent,
as the lower example in Figure~\ref{fig:intro}(c) shows. Admission must
track readiness at each level of sharing, delaying dependent requests
while allowing independent requests to run.

\vspace{2mm}

\noindent\textbf{Contributions.}
To address these challenges, we propose \textbf{KaleidoKV}, a framework that
plans evidence orders across an offline batch and controls when
requests enter the engine to reuse shared prefixes. Figure~\ref{fig:intro}(d) shows how KaleidoKV uses the layout to
evaluate grouping changes and track dependencies during execution.

\vspace{0.5mm}

\noindent\raisebox{0.2ex}{\scriptsize$\bullet$}\enspace
\textbf{Joint layout planning.}
To improve sharing beyond
early grouping decisions, KaleidoKV revises formed groups through joint
moves of several requests. We prove that minimizing planned prefill is
NP-hard even when every request retrieves two evidence items. Therefore, KaleidoKV solves
small scopes exactly and refines larger scopes through a bounded
number of regrouping rounds. We also show that moving
several requests together can save more than moving each of them alone
only if the move creates or empties a prefix that several requests
share. These two cases lead to two moves: a merge gathers requests from
different groups onto a new shared prefix, and a dissolve removes a
group whose members reuse more in other groups than the group saves.
KaleidoKV keeps a merge or dissolve only if it lowers the planned
prefill of the whole batch.

\vspace{0.5mm}

\noindent\raisebox{0.2ex}{\scriptsize$\bullet$}\enspace
\textbf{Incremental evaluation on a global prefix index.}
To keep joint refinement efficient, KaleidoKV evaluates proposed
changes incrementally without scanning the whole batch. A global
prefix index records how many requests use each prefix and computes
the exact change in planned prefill from only the old and new paths
of the moved requests, in $O(k)$ time per moved request with at most
$k$ evidence items. We further prove that, with the orders of other requests fixed, a best order for a request can be found by considering at most \(2^k\) representative orders, rather than all \(k!\) permutations considered in prior
work~\cite{liu2025optimizing, li2026solo}.

\vspace{0.5mm}

\noindent\raisebox{0.2ex}{\scriptsize$\bullet$}\enspace
\textbf{Prefix-ready admission.}
To realize planned sharing during concurrent execution, KaleidoKV
admits requests according to shared-prefix readiness. It tracks
cached, in-flight, and pending
segments, delaying requests that would duplicate ongoing computation
while admitting requests that can compute independent prefixes. For
nested groups, each request waits for the shared segments on its path
that other requests are still computing.

\vspace{0.5mm}

\noindent\raisebox{0.2ex}{\scriptsize$\bullet$}\enspace
\textbf{End-to-end efficiency, prefix reuse, and scalability.}
\textbf{(a)} KaleidoKV achieves the highest throughput and lowest mean
TTFT in all 16 settings across four QA workloads and four models,
with approximately 1.24--1.75$\times$ the throughput
of the strongest baseline, including planning time.
\textbf{(b)} With Llama-3.1-8B under the default configuration,
KaleidoKV reduces actual prefill by 37.0--61.5\% across the four
workloads relative to SGLang (DFS) \cite{zheng2024sglang}, which retains the original evidence
order and submits requests in depth-first order of their prefix tree.
KaleidoKV keeps actual prefill within 0.5\% of planned prefill.
\textbf{(c)} On WikiCities, scaling the batch from 1{,}000 to
8{,}000 requests increases planning time from 4.9 to 23.3 seconds,
compared with 3.8 to 165.6 seconds for ContextPilot \cite{jiang2026contextpilot}.
\par
\endgroup

\vspace{-2mm}

\section{Related Work} \label{sec:related_work}
\vspace{-1mm}


\noindent \textbf{Prefix Caching in RAG.} In RAG, a prompt contains a fixed instruction, the passages
retrieved for a question, which we call evidence items, and the
question itself. The evidence items are usually the longest part of
the prompt, so most of the prefill cost is spent on them. Prefix
caching keeps the KV cache of earlier prompts and reuses it when a new
prompt starts with the same tokens~\cite{kwon2023vllm, zheng2024sglang}.
PolarKV~\cite{zhang2026polarkv} extends
cache capacity through paired cloud-memory and storage shards,
keeping most tiering traffic local.
Since the KV of a token depends on all tokens before it,
the reused part must match token by token from the start of the
prompt. 
Prefix caching saves evidence prefill only when requests place identical evidence in the same order, and while cached states remain in memory. Existing work improves KV cache reuse for RAG along three directions, including approximate evidence reuse at any position, exact reuse of shared prefixes, and scheduling for prefix reuse across requests.


\noindent \textbf{Approximate KV Cache Reuse.}
Retrieval places the same evidence item at
different positions across prompts. Several methods reuse independently
computed KV states beyond shared prefixes, correcting errors from
missing attention to preceding items. Selective recomputation limits
this correction to a subset of tokens.
CacheBlend~\cite{yao2025cacheblend} selects the tokens
whose KV states deviate most from those of a full prefill,
EPIC~\cite{hu2025epic} recomputes the first few tokens of each item,
and Cache-Craft~\cite{agarwal2025cache} also decides which items to
keep in its cache. APE~\cite{yang2025ape} instead adjusts the attention
over the separately encoded items without recomputation. Others
fine-tune the model to accept such items, through position
re-encoding~\cite{ma2025block} or trainable link
tokens~\cite{yang2026kvlink}. However,
these methods retain approximation error or require model adaptation,
leaving exact reuse across prompts under unchanged attention semantics unresolved.


\noindent \textbf{Exact Prefix Reuse.} Exact prefix reuse gives the
same KV states as a full prefill of the prompt, and existing methods
increase it by creating more shared prefixes. They fall into two
categories, which change the evidence of the requests or only its
order. The first category gives related requests a common evidence
context and computes its KV states once.
SubGCache~\cite{zhu2026subgcache} clusters the queries of graph-based
RAG and builds one representative subgraph for each cluster, and
cache-augmented generation~\cite{chan2025don} preloads a whole document
collection instead of retrieving from it. Each request then attends to
evidence that it did not retrieve, which adds computation and may lower
answer quality. The second category keeps the evidence of each request
and reorders it into shared prefixes.
ContextPilot~\cite{jiang2026contextpilot} clusters requests by the
distance between their evidence and aligns the evidence within each
cluster, while GGR~\cite{liu2025optimizing} and SOLO~\cite{li2026solo}
reorder the fields and rows of tables with greedy algorithms.
However, fixed greedy groups can miss better
sharing across overlapping evidence sets, while pairwise comparisons
and repeated scans increase planning cost. Efficient regrouping
without changing retrieved evidence remains unresolved in large batches.

\noindent \textbf{Scheduling for Prefix Reuse.}
Scheduling determines whether shared prefixes
remain cached until reuse. Serving systems exploit this locality
through cache-aware request ordering.
SGLang~\cite{zheng2024sglang} keeps the cached prefixes
in a radix tree, evicts the least recently used ones, and can serve the
waiting requests with the longest cached prefix first.
Preble~\cite{srivatsa2025preble} extends prefix-aware scheduling to
several GPUs and balances prefix reuse against load, and Dexter et
al.~\cite{dexter2026llm} order requests that share prefixes under
latency constraints. For offline batches,
BatchLLM~\cite{zheng2026batchllm} extracts the common prefixes of a
batch and computes each of them once before the requests that use them,
and BlendServe~\cite{zhao2026blendserve} orders requests on a prefix
tree so that compute-intensive and memory-intensive requests overlap
while their prefixes remain shared.
HotPrefix~\cite{li2025hotprefix} manages
CPU--GPU cache residency by tracking prefix hotness.
KVFlow~\cite{pan2025kvflow} uses agent execution graphs to guide
eviction and prefetching, delaying requests while their KV states
are loading. These methods schedule requests in the given prompt
orders to reuse existing shared prefixes, leaving the shared-prefix
computation dependencies introduced by evidence rearrangement across
requests in a batch unaddressed.

\begin{table}[!t]
\caption{Frequently used notation}
\label{tab:notation}
\footnotesize
\centering
\setlength{\tabcolsep}{3pt}
\setlength{\arrayrulewidth}{0.4pt}
\renewcommand{\arraystretch}{1.04}
\begin{tabularx}{\columnwidth}{
    |>{\raggedright\arraybackslash}p{0.20\columnwidth}|
    >{\raggedright\arraybackslash}X|
}
\hline
\rowcolor{black!25}
\textbf{Notation} & \textbf{Description} \\ \hline
$\mathcal{R}$
    & The set of input requests \\ \hline
$\mathcal{E}$
    & The set of distinct evidence items retrieved by $\mathcal{R}$ \\ \hline
$\mathcal{E}(r)$
    & The evidence set retrieved by request $r$ \\ \hline
$\ell(e)$
    & The token length of evidence item $e$ \\ \hline
$O = \{o_r\}$
    & The layout, where $o_r$ is the evidence order of request $r$ \\ \hline
$W$, $W'$
    & The planned prefill before and after a layout change \\ \hline
$\Delta W$
    & The reduction in planned prefill, $W-W'$ \\ \hline
$G$, $p_G$
    & A request group and its longest shared prefix \\ \hline
$u_r$, $u_r'$
    & The reuse of request $r$ before and after a layout change \\ \hline
$\sigma(G)$
    & The prefill saving contributed by group $G$ \\ \hline
$n_v$
    & The number of requests passing through prefix-tree node $v$ \\ \hline
$L(X)$
    & The list of requests whose evidence sets contain $X$ \\ \hline
\end{tabularx}
\end{table}

\begin{figure*}[t]
  \centering
  \includegraphics[width=\textwidth]{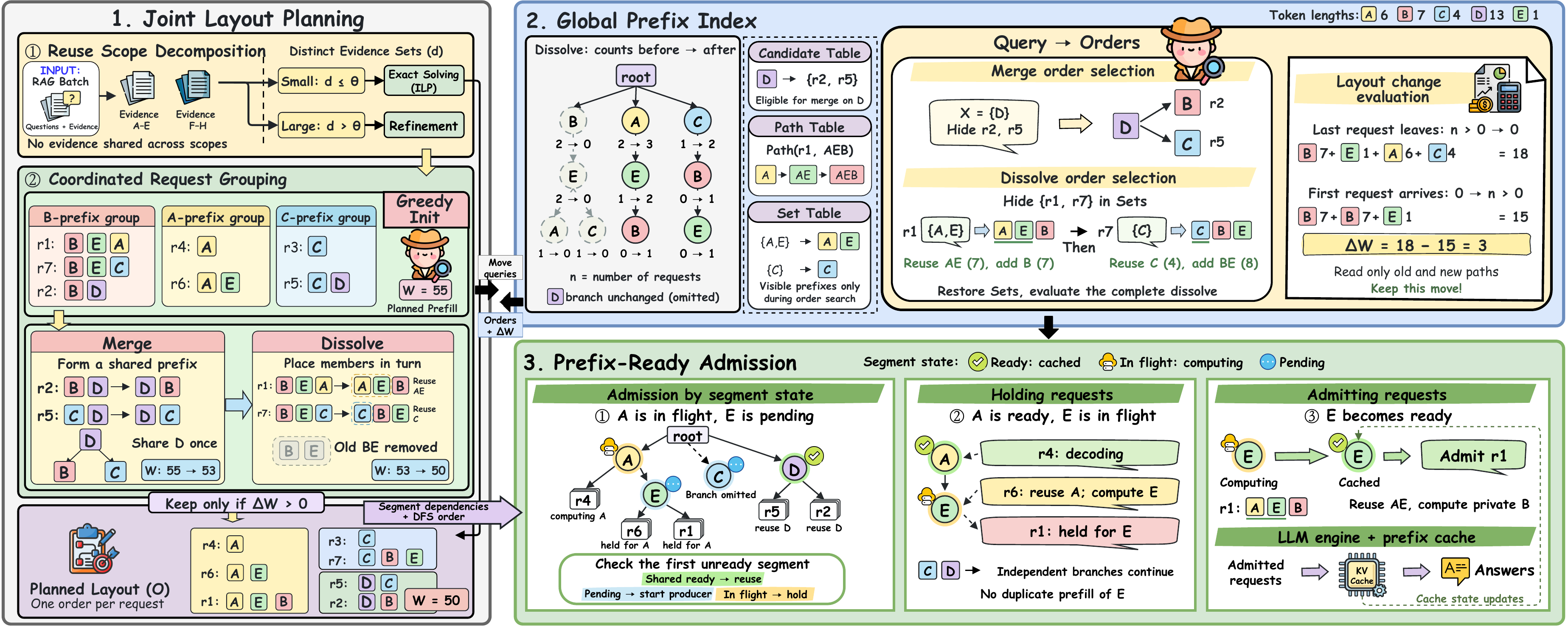}
  \caption{Overview of \sysname{}. (1) Joint layout planning rearranges each request's evidence to expose shared prefixes. (2) The Global Prefix Index selects candidate orders and evaluates their prefill savings. (3) Prefix-ready admission uses the planned dependencies and runtime cache states to hold dependent requests and admit them when their shared prefixes become ready.}
  \vspace{2mm}
  \label{fig:framework}
\end{figure*}

\vspace{-2mm}

\begingroup
\section{Preliminary}
\label{sec:preliminary}

\vspace{-1mm}

\nopagebreak[4]

\myparagraph{Workload and evidence}
Given an offline RAG batch of $n$ requests $\mathcal{R}=\{r_1,\dots,r_n\}$,
all questions and retrieved evidence are known in advance.
Let $\mathcal{E}=\{e_1,\dots,e_m\}$ be its distinct, fixed-text evidence items.
A request $r$ consists of a question $q(r)$
and its evidence set $\mathcal{E}(r)\subseteq\mathcal{E}$.
For a nonempty request set $A\subseteq\mathcal{R}$,
$\mathcal{E}(A)=\bigcup_{r\in A}\mathcal{E}(r)$ and
$I(A)=\bigcap_{r\in A}\mathcal{E}(r)$ denote the union and intersection
of its evidence sets. Table~\ref{tab:notation} summarizes notations used in this paper.

\begin{definition}[Layout]
\label{def:layout}
A layout $O=\{o_r\}_{r\in\mathcal{R}}$ assigns each request $r$
an order $o_r$ of $\mathcal{E}(r)$, with each item appearing exactly once.
The LLM receives this ordered evidence together with $q(r)$ in its prompt.
\end{definition}

\noindent
Since KV states depend on preceding tokens, exact reuse requires
matching token prefixes. At evidence-item boundaries, a layout forms
a prefix tree. Its empty root has zero weight, and each non-root node
holds one evidence item. Requests follow their orders from the root,
sharing nodes only along identical initial sequences.

\begin{definition}[Planned prefill]
\label{def:planned-prefill}
Let $\ell(e)>0$ be item $e$'s token length. Planned prefill is
$W=\sum_v\ell(e_v)$, where $v$ ranges over used non-root nodes
holding item $e_v$. Each node counts once even when shared.
Instruction and question tokens are excluded from the weight $W$.
\end{definition}

\noindent
Actual execution may recompute shared prefixes whose KV
states are unavailable. Reordering evidence changes the tree weight
from $W$ to $W'$, a planned prefill reduction of $\Delta W=W-W'$.

\begin{definition}[Request reuse]
\label{def:request-reuse}
For a node $v$, let $n_v$ be the number of requests passing through
it. The potential reuse $u_r$ of request $r$ is the total token length
of its path nodes also traversed by other requests. We write $u'_r$
for this quantity on its new path after a layout change.
\end{definition}

\begin{definition}[Request group]
\label{def:group}
A group $G$ consists of all requests traversing a non-root node
with $n_v\ge1$. Nodes with identical request sets form one group.
Its longest shared prefix is $p_G$, and its nodes are those on $p_G$
traversed by no request outside $G$ in the current layout.
\end{definition}

\noindent
The saving $\sigma(G)$ is $|G|-1$ times the total weight of $G$'s
nodes and is zero for singletons. Subtracting all distinct groups'
savings from the requests' total evidence length gives the layout weight $W$.

\myparagraph{Problem statement}
Given $\mathcal{R}$ and $\mathcal{E}(r)$ for every request $r$, the
task is to choose a layout and schedule minimizing the time from
the start of processing until all answers are generated, including
planning and scheduling time. Each order must retain exactly the
retrieved evidence items and preserve their content. Since evidence
prefill dominates inference cost, planning aims to lower $W$ within
affordable time, while scheduling seeks to realize this prefix sharing.

\par
\endgroup

\vspace{5mm}
\section{Method} 
\vspace{1mm}
\label{sec:method}

\noindent \textbf{Overview.}
To reduce batch completion time, KaleidoKV combines evidence-order
planning with request admission. Given the requests and their retrieved
evidence, it builds a prefix-tree layout to expose reuse across the
batch, using an index to search candidate orders and evaluate layout
changes efficiently. During execution, admission uses the planned
sharing dependencies and the engine's prefix cache to decide when each
request can enter. Each request retains its evidence set and content, and the engine's
attention kernels remain unchanged.
Figure~\ref{fig:framework} outlines the following three components:

\vspace{1mm}
\noindent\raisebox{0.2ex}{\scriptsize$\bullet$}\enspace \textbf{Joint Layout Planning.}
This component constructs an optimized layout by jointly planning the evidence orders of requests. It splits the workload into independent reuse scopes and solves small scopes exactly. For large scopes, coordinated request grouping refines a greedy layout through two operations: a merge gathers requests under a new shared prefix, while a dissolve redistributes the members of an existing group. Each change is accepted only if it reduces the planned prefill of the whole workload (Section~\ref{sec:grouping}).

\vspace{1mm}

\noindent\raisebox{0.2ex}{\scriptsize$\bullet$}\enspace \textbf{Global Prefix Index.}
During grouping, the index finds orders and evaluates layout changes. Reference counts give the exact cost change from the moved
requests' paths. Candidate lists organize merges by evidence set, and
a set table locates reusable prefixes without enumerating their
permutations. For each merge or dissolve placement, it selects an order before evaluating its gain
(Section~\ref{sec:index}).

\vspace{1mm}

\noindent\raisebox{0.2ex}{\scriptsize$\bullet$}\enspace \textbf{Prefix-Ready Admission.}
Given the planned layout, this component controls request admission
according to prefix readiness. Requests are considered in the layout's depth-first
order to keep members of each group close together. It tracks whether each part of a shared prefix is cached, being
computed, or not yet started. It holds requests whose shared prefix is
being computed and admits requests from other branches in the meantime.
Waiting requests are released as the shared prefixes they need become
ready
(Section~\ref{sec:admission}).

\vspace{5mm}

\subsection{Joint Layout Planning}
\label{sec:grouping}
\vspace{1mm}

Joint layout planning improves prefix sharing by jointly planning
the evidence orders of requests while preserving each request's
evidence set. Since each request has one evidence
order, joining one group can sacrifice reuse with another. Therefore, the planner
evaluates each layout change by its effect on the weight $W$
of the whole prefix tree and accepts it only if $W$ decreases. Some
improvements require several requests to move together, even when each
move alone raises $W$, as Figure~\ref{fig:intro} illustrates. Finding the best combination of evidence orders is NP-hard,
which motivates a scalable planner that efficiently reduces
planned prefill with bounded search effort.

\begin{theorem}[Hardness]
\label{thm:hardness}
Choosing the evidence orders of all requests so that their prefix tree
has the smallest weight is NP-hard.
\end{theorem}

\noindent
Theorem~\ref{thm:hardness} follows by reduction from minimum
vertex cover~\cite{karp1972reducibility}. Appendix A of the online supplementary
material~\cite{appendix} provides the full reduction and proofs of all other omitted
theoretical results.

\vspace{1mm}
\noindent\textbf{Reuse scope decomposition.}
Reuse scopes separate the workload into pieces
whose evidence orders can be planned independently. Requests in
different scopes share no evidence, so planning each scope independently
preserves all evidence reuse across the batch.

\begin{definition}[Reuse scope]
\label{def:scope}
Given the request set $\mathcal{R}$ and the evidence set
$\mathcal{E}$, a \emph{reuse scope} is a maximal request set
$S\subseteq\mathcal{R}$ such that any division of $S$ into two
nonempty parts $A$ and $B$ satisfies
$\mathcal{E}(A)\cap\mathcal{E}(B)\neq\emptyset$. The set of all
scopes, denoted by $\mathcal{S}$, partitions $\mathcal{R}$ and
satisfies two properties:
(i) $\bigcup_{S\in\mathcal{S}}\mathcal{E}(S)=\mathcal{E}$, and
(ii) $\mathcal{E}(S)\cap\mathcal{E}(S')=\emptyset$ for any two scopes
$S\neq S'$, so no evidence prefix crosses scope boundaries.
\end{definition}

\noindent
Lemma~\ref{lem:scopes} shows that
this split loses nothing, so finding the minimum weight can be reduced
to finding it within each scope.

\begin{lemma}[Scope decomposition]
\label{lem:scopes}
Given the request set $\mathcal{R}$ and its scope set $\mathcal{S}$,
placing a request of a scope $S\in\mathcal{S}$ with the requests of
another scope $S'\in\mathcal{S}$ can never lower the prefix tree
weight of $\mathcal{R}$.
\end{lemma}

\noindent
Every group lies within one scope because its members share evidence.
We plan each scope in turn to lower $W$ while limiting planning cost. Scopes with few distinct evidence sets admit low-cost exact
solutions, whereas larger scopes need refinement.

\noindent \uline{Small-scope optimization.}
We call a scope small if it has at most $\theta$ distinct
evidence sets, and we solve a small scope $S$ exactly with integer
linear programming (ILP)~\cite{nemhauser1988integer}. For a request $r\in S$, let
$\mathcal{O}(r)$ denote the orders of its evidence items. For an order $o$, let $\mathcal{P}(o)$ denote its nonempty 
prefixes, and $\mathcal{P}$ be the union of $\mathcal{P}(o)$ over
all orders of requests in $S$. Binary variables $x_{r,o}$ and $y_t$ equal $1$
if $r$ takes $o$ and if prefix $t$ is computed, respectively. The ILP is
\begin{equation}
\begin{aligned}
\min\ \ & \textstyle\sum_{t\in\mathcal{P}}\ell(e_t)\,y_t\\
\text{s.t.}\ \ & \textstyle\sum_{o\in\mathcal{O}(r)}x_{r,o}=1,\quad \forall r\in S,\\
& x_{r,o}\le y_t,\quad \forall r\in S,\, o\in\mathcal{O}(r),\, t\in\mathcal{P}(o),\\
& x_{r,o},\,y_t\in\{0,1\},
\end{aligned}
\label{eq:scope_ip}
\end{equation}
where $e_t$ is the last evidence item of $t$. The first constraint
gives every request one order, and the second, as in facility
location~\cite{nemhauser1988integer}, requires every prefix of a taken
order to be computed. The minimization sets $y_t=0$ for all other
prefixes, so the objective is the weight of the prefix tree of $S$, in
which a shared prefix is paid once. In Figure~\ref{fig:intro}(d), an optimal solution gives $r_3$ the order
$(A,B,D)$ and $r_5$ $(C,D)$, so the shared prefixes $(A,B)$
and $(C)$ are each paid once, for the minimum weight of 30. Requests with the same evidence set can take the same order without
raising $W$, so the ILP needs only one order choice per distinct
evidence set. A dynamic program reaches the
same minimum without a solver (Appendix B \cite{appendix}).

\noindent\textbf{Coordinated request grouping.}
A large scope may contain thousands of distinct
evidence sets, making exact solving too costly. We refine its layout
by moving requests between groups and first identify when joint moves
can improve on individual ones. For a single-request move of
$r$, the reduction of $W$ equals $u_r'-u_r$. Lemma~\ref{lem:joint} identifies when joint gain exceeds the sum of individual gains.

\begin{lemma}[Sources of joint gain]
\label{lem:joint}
For a change that moves a set $M$ of requests together, let
$\Delta W(M)$ be its reduction of $W$, and let $\Delta W(r)$ be the
reduction when $r\in M$ moves alone to the same order. Then
$\Delta W(M)>\sum_{r\in M}\Delta W(r)$ only if the change (i) creates a
prefix that no request used before and at least two requests of $M$
share, or (ii) empties a prefix that at least two requests shared
before.
\end{lemma}

\noindent
Lemma~\ref{lem:joint} motivates revising whole groups in large scopes, starting from a greedy layout and performing joint request moves.

\vspace{1mm}
\noindent\uline{Greedy layout initialization.}
Joint regrouping starts from a greedy grouping~\cite{liu2025optimizing},
which in each step groups all requests that retrieve the evidence whose
sharing saves the most prefill tokens, and never revisits a formed group.
Groups are defined on the nodes of one tree, so two groups are either
nested or disjoint, and a group may contain smaller groups. Lemma~\ref{lem:joint} suggests two moves that
revise these groups, each changing a group as a whole, namely forming a
new group and removing an existing one. We apply them in two steps,
merges for case (i) and dissolves for case (ii). Since both steps form or
remove groups, we evaluate their moves by the savings of the groups. By
Section~\ref{sec:preliminary}, $W$ equals the total evidence length minus
the savings of all groups, so lowering $W$ means raising their total
saving. A merge creates the saving of a new group but lowers the savings
of the groups that its requests leave. A dissolve removes a group, and it
is kept only if its members reuse more in other groups than the group
saves. A dissolve places members sequentially against the prefixes available
outside the old group, including those created by earlier placements.
It does not enumerate joint destinations for all members.
Dissolve follows merge so that its members can reuse the newly formed
groups when choosing their new evidence orders.

\begin{figure}[t]
    \centering
    \includegraphics[width=.9\linewidth]{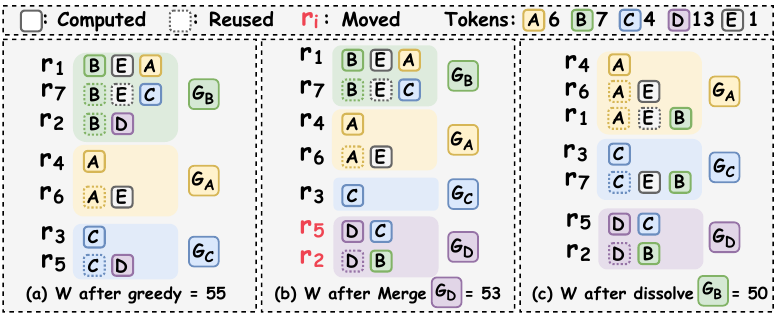}
    \caption{Merge and dissolve on one scope.}
    \label{fig:regroup}
\end{figure}

\vspace{1mm}
\noindent\uline{Merge: forming groups on new prefixes.}
A merge gathers requests under a common prefix.
Enumerating all combinations of current groups would be exponential
in the number of groups. Instead, we use the evidence at the proposed
prefix to identify eligible requests. For a nonempty evidence set $A$,
let $L(A)=\{r\mid A\subseteq\mathcal{E}(r)\}$ and
$\ell(A)=\sum_{e\in A}\ell(e)$. All orders $t$ of $A$ have the same
eligible requests and length, so $L(t)=L(A)$ and $\ell(t)=\ell(A)$.
A merge chooses $M\subseteq L(A)$ and moves its members to orders
starting with $t$, preserving the current relative order of their
remaining items. For an unused prefix, Lemma~\ref{lem:merge} relates group savings to members' lost reuse.

\begin{lemma}[Gain of a merge]
\label{lem:merge}
Let $t$ be a sequence of distinct evidence items such that no request
uses any nonempty prefix of $t$ yet, and let a merge move a set $M\subseteq L(t)$
of requests onto $t$. If every non-singleton group that loses requests keeps other
members and the requests of $M$ share only the prefix $t$, the merge
lowers $W$ by
\begin{equation}
\Delta W=\sum_{r\in M}\bigl(\ell(t)-u_r\bigr)-\ell(t)
=\sigma(G_t)-\sum_{r\in M}u_r,
\label{eq:merge}
\end{equation}
where $\sigma(G_t)=(|M|-1)\,\ell(t)$ is the saving of the new group
$G_t$ on $t$. Hence the merge lowers $W$ if and only if $G_t$ saves
more than the requests of $M$ reuse in the current layout, which
requires at least two of them to reuse fewer than $\ell(t)$ tokens in the layout before the merge.
\end{lemma}

\noindent
The saving $\sigma(G_t)$ in Eq.~\eqref{eq:merge} is at
most $(|L(A)|-1)\ell(A)$. This priority depends only on the evidence
set and the workload. Each request contributes at most $2^k-1$
nonempty subsets, so a workload of $N$ requests has at most
$(2^k-1)N$ distinct candidate sets. We order the sets with
$|L(A)|\ge2$ once by decreasing priority and reuse this order in
every merge round.
For each $A$, we first try moving all of $L(A)$.
We then partition $L(A)$ by each request's current first evidence
item and try each part once, keeping this partition fixed during
these attempts. Members of a part retain their shared first item:
it either belongs to $A$ or becomes the first item after $t$.
For each move group $M$, the index selects the order $t$ of $A$
that minimizes the resulting weight (Section~\ref{sec:index}).
The move is accepted only when its exact gain is positive.
Once $t$ is in use, a request joining it can reuse at least
$\ell(t)$ tokens, and its gain is $u'_r-u_r\ge\ell(t)-u_r$.
Each round ends with single-request improvements.
We run at most $I_{\max}$ rounds and stop when a round keeps no
change to the layout.

\begin{example}
In Figure~\ref{fig:regroup}(a), no request currently starts with $D$, and
$L(D)=\{r_2,r_5\}$. These two requests currently reuse $u_{r_2}=7$ and
$u_{r_5}=4$ tokens, both fewer than $\ell(D)=13$. By
Eq.~\eqref{eq:merge}, merging them onto $D$ lowers $W$ by
$(13-7)+(13-4)-13=2$, from 55 to 53, whereas moving $r_2$ or $r_5$
alone raises $W$ to 62 or 59, respectively. The merge forms the group
$G_D$ of Figure~\ref{fig:regroup}(b) and removes $r_2$ from $G_B$.
\end{example}

\noindent\uline{Dissolve: removing premature groups.}
After merging, a group may retain fewer members or
save less than its members could reuse elsewhere. A dissolve removes
such a group $G$ as a whole. Its members leave one at a time. Each
chooses the order with the greatest gain that avoids the nodes of
$G$, taking earlier members' new orders into account. We call $G$
premature if its completed dissolve lowers $W$. Here, $u'_r$ is the
reuse of $r$ immediately after its move.
Lemma~\ref{lem:dissolve} gives the gain of the complete dissolve under the stated conditions.

\begin{lemma}[Gain of a dissolve]
\label{lem:dissolve}
Let no request outside a group $G$ use any nonempty prefix of $p_G$, and no
two members of $G$ share evidence items after $p_G$. While another member
stays in $G$, moving a member $r$ out lowers $W$ by $u_r'-\ell(p_G)$,
and moving all out lowers $W$ by
\begin{equation}
\Delta W=\ell(p_G)-\sum_{r\in G}\bigl(\ell(p_G)-u_r'\bigr)
{}=\sum_{r\in G}u_r'-\sigma(G),
\label{eq:dissolve}
\end{equation}
where $\sigma(G)=(|G|-1)\,\ell(p_G)$. Hence $G$ is premature if and only
if its members reuse more in other groups than $G$ saves.
\end{lemma}

\noindent
For $\left|G\right| \ge 2,$ each $r$ in Eq.~\eqref{eq:dissolve}  computes
$\ell(p_G)-u_r'$ more tokens at its new place, and once the last member
leaves, $p_G$ is no longer computed, which returns $\ell(p_G)$ tokens
as the joint gain of case (ii) in Lemma~\ref{lem:joint}. Unlike a merge,
which pays $\ell(t)$ once and must recover this
cost through the combined gains of its requests, a dissolve lowers $W$ even when every member
computes more, as long as their extra computation stays below $\ell(p_G)$.
Moving only some members cannot earn this return, so a dissolve
evaluates the move of the whole group. Lemma~\ref{lem:dissolve} covers
a group that lies in no larger group. When a nested group $G$ is dissolved,
its members leave only the nodes of $G$ and may keep the prefix of the
larger group, and the index still evaluates this move exactly
(Section~\ref{sec:index}). Each round visits the groups present at its start once, skipping
those that have since gained members. A request can be
placed at most $k$ times in a round, once for each group on its
initial path. Within
a dissolve, later members can share the evidence that earlier ones
place, and requests outside the group keep their orders. The rounds
repeat until a round dissolves no group or $I_{\max}$ rounds have run.

\begin{example}
In Figure~\ref{fig:regroup}(b), $r_1$ and $r_7$ remain in $G_B$ and
share the prefix $(B,E)$ of 8 tokens. After leaving $G_B$, $r_1$ reuses
$A$ and $E$ in $G_A$ and $r_7$ reuses $C$ with $r_3$, so
$u_{r_1}'=7$ and $u_{r_7}'=4$. By Eq.~\eqref{eq:dissolve}, dissolving
$G_B$ lowers $W$ by $8-\bigl((8-7)+(8-4)\bigr)=3$, from 53 to 50,
whereas moving $r_1$ or $r_7$ out alone raises $W$ to 54 or 57,
respectively. The two members join different groups, so no single merge realizes these two destination prefixes. At the greedy start, where
$r_2$ also shared $B$ and had no group to join, the same dissolve would
raise $W$ by 4, so $G_B$ became premature only after the merge onto $D$.
\end{example}

\begin{algorithm}[t]
	{
		\SetVline
		\footnotesize
		\caption{Joint Layout Planning}
		\label{alg:grouping}
	\SetCommentSty{textbf}
	\KwIn{requests $\mathcal{R}$, evidence $\mathcal{E}$, threshold $\theta$ on distinct evidence sets per small scope, round limit $I_{\max}$ per phase}
	\KwOut{a layout $O=\{o_r\}$}
	\State{$\mathcal{S}\leftarrow\textnormal{ReuseScopes}(\mathcal{R})$}
	\State{$O\leftarrow\textnormal{GreedyGrouping}(\mathcal{R})$}
	\StateCmt{$\textnormal{BuildIndex}(O)$}{Algorithm~\ref{alg:index}}
	\For{$S\in\mathcal{S}$}{
		\uIf(\tcc*[f]{Small scope}){$|\{\mathcal{E}(r):r\in S\}|\le\theta$}{
			\StateCmt{$\mathit{change}\leftarrow\textnormal{SolveILP}(S)$}{Eq.~\eqref{eq:scope_ip}}
			\lIf{$\textnormal{ComputeDeltaWeight}(\mathit{change})>0$}{$\textnormal{ApplyChange}(O,\mathit{change})$\;}
		}
		\Else(\tcc*[f]{Large scope}){
			\State{$\textnormal{Merge}(S)$}
			\For(\tcc*[f]{Dissolves}){$i\leftarrow1$ \KwTo $I_{\max}$}{
				\State{$W_0\leftarrow W$}
				\State{$\mathcal{G}_0\leftarrow\textnormal{SnapshotGroups}(S)$}
				\For{$G\in\mathcal{G}_0$}{
					\lIf{$\textnormal{GainedMembers}(G)$}{\textbf{continue}\;}
					\State{$(\mathit{change},\Delta W_G)\leftarrow\textnormal{PlanForDissolve}(G)$}
					\lIf{$\Delta W_G>0$}{$\textnormal{ApplyChange}(O,\mathit{change})$\;}
				}
				\lIf{$W=W_0$}{\textbf{break}\;}
			}
		}
	}
	\State{\Return $O$}
\vspace{6.3pt}
\textbf{Procedure} $\textnormal{Merge}(S)$\\
\Indp

\For{$i\leftarrow1$ \KwTo $I_{\max}$}{
    \State{$W_0\leftarrow W$}
    \For{$X\in\textnormal{RankCandidates}(S)$}{
        \State{$t\leftarrow\textnormal{OrderForMerge}(X,L(X))$}
        \StateCmt{$\textnormal{TryMerge}(L(X),t)$}{Accept only if $\Delta W>0$}
        \State{$\mathcal{P}\leftarrow\textnormal{PartitionByFirst}(L(X),O)$}
        \For{$M\in\mathcal{P}$}{
            \State{$t\leftarrow\textnormal{OrderForMerge}(X,M)$}
            \State{$\textnormal{TryMerge}(M,t)$}
        }
    }
    \lFor{$r\in S$}{$\textnormal{TryImprove}(r)$\;}
    \lIf{$W=W_0$}{\textbf{break}\;}
}
\State{\Return}
\Indm
	}
\end{algorithm}

\noindent\textbf{Planning cost analysis.}
Without an index, selecting an order can enumerate
all permutations, and evaluating a candidate by rebuilding the
prefix tree takes $O(kN)$ time. Theorem~\ref{thm:cost_plain} bounds
this baseline for comparison with indexed search
(Section~\ref{sec:index}).

\begin{theorem}[Planning cost]
\label{thm:cost_plain}
Let the workload hold $N$ requests, each retrieving at most $k$
evidence items, and let merges and dissolves each run at most
$I_{\max}$ rounds. If candidate orders are enumerated
explicitly and every change is evaluated by recomputing $W$,
the scope split, the greedy start, and the merges and dissolves take
$O(I_{\max}\cdot k!\cdot k^2\cdot N^2)$ expected time in total, apart from
the ILP solves.
\end{theorem}

\noindent\textbf{Algorithm description.}
Algorithm~\ref{alg:grouping} initializes the scopes, greedy layout,
and index (lines 1--3), then accepts improving ILP solutions for small
scopes (lines 5--7). For large scopes, $\textnormal{RankCandidates}$ caches
nonempty sets $X$ with $|L(X)|\ge2$ by decreasing
$(|L(X)|-1)\ell(X)$ (line 23). $\textnormal{OrderForMerge}$ selects a prefix
for all of $L(X)$ or one part (lines 24 and 28).
$\textnormal{PartitionByFirst}$ snapshots the current first-item partition
after the full-group attempt (line 26). $\textnormal{TryMerge}$ accepts a
joint move only if $\Delta W>0$, preserving residual evidence order
(lines 25 and 29). $\textnormal{TryImprove}$ similarly tests each request's
best order (line 30). For dissolves, $\textnormal{SnapshotGroups}$ records
live group handles and tracks committed arrivals. $\textnormal{GainedMembers}$
skips any group that subsequently gained members (lines 12--14).
$\textnormal{PlanForDissolve}$ proposes a complete joint move, accepted only
with positive gain (lines 15--16). Both phases stop after $I_{\max}$
rounds or a round with unchanged planned prefill $W$ (lines 17 and 31).
\vspace{-1mm}

\subsection{Global Prefix Index}
\label{sec:index}
\vspace{-1mm}

Merges and dissolves try many changes, and each try
must choose an order and evaluate it. The index keeps both steps cheap.
It finds one order by looking up the evidence that an order can reuse,
instead of trying $k!$ orders, and it evaluates this order from the
moved paths, instead of rebuilding the tree after every trial move.

\noindent\textbf{Prefix reference counting.}
We maintain a prefix tree over evidence-item sequences, similar to those used in LLM serving~\cite{zheng2024sglang,
zheng2026batchllm, jiang2026contextpilot}. Each node \(v\) holds one evidence item
$e_v$ of $\ell(e_v)$ tokens. Each evidence item is tokenized on its
own, and we check that tokenizing a whole prompt gives the same tokens,
so a shared path of nodes is a shared token prefix.
The tree starts from the current layout, and a proposed
path is inserted on first use with zero counts at its new nodes. Each
node keeps the number $n_v$ of requests whose current orders pass
through it, giving $W=\sum_{v:\,n_v>0}\ell(e_v)$.

\noindent\textbf{Index table organization.}
Three tables answer the questions that searches and
evaluations ask. The \textit{candidate table} tells which requests can
share a prefix. For each evidence set $X$ that at least two requests
retrieve, it stores the list $L(X)$. The \textit{path table} tells which
nodes a request passes under a given order. It stores the nodes of each
order that planning has proposed, so an evaluation reads them directly
instead of following the tree from the root. The \textit{set table}
tells which evidence a moved request can reuse. Its entry
$\mathit{Sets}[X]$ lists the nodes whose root paths hold exactly the
items of $X$, such as the last nodes of $(A,B)$ and $(B,A)$ for
$\{A,B\}$, and its entry $\mathit{Sets}[X,a]$ lists the children with
item $a$ of these nodes. The root holds the empty set and is always
listed as the zero-reuse fallback.

\noindent\textbf{Layout change evaluation.}
A change gives some requests new orders. Lemma~\ref{lem:crossing} shows
that its $\Delta W$ depends only on the nodes that lose their last
request or gain their first one under the new layout.

\begin{lemma}[Evaluation by reference counts]
\label{lem:crossing}
For a change that gives some requests new orders, let $V^-$ be the
nodes that lose their last request and $V^+$ the nodes that gain their
first one. Then
\begin{equation}
\Delta W=\sum_{v\in V^-}\ell(e_v)-\sum_{v\in V^+}\ell(e_v),
\label{eq:crossing}
\end{equation}
where both sets lie on the old and new paths of these requests.
\end{lemma}

\noindent
Therefore, an evaluation of $\Delta W$ reads only the old and new paths
and changes nothing. An accepted change updates the counts and the set
table on these paths and subtracts $\Delta W$ from $W$.

\noindent\textbf{Merge order selection.}
A merge moves a group $M\subseteq L(X)$ onto an order
$t$ of an evidence set $X$, and each member keeps the rest of its order.
Choosing $t$ by enumerating all $|X|!$ orders would require evaluating each on every member of $M$. The index instead asks which order reuses
the most tokens that other requests already compute.
To answer this, the index hides $M$ in the set table, so
that the visible nodes are those of the other requests. Removing the
items of $X$ from the members' orders leaves a residual tree $U_M$,
whose shape does not depend on $t$. The new paths are $t$ followed by
$U_M$, so their total weight is $\ell(X)+W(U_M)$ for every $t$. Therefore, the $t$
that gives the smallest $W$ is the one whose new paths reuse
the most visible tokens. Lemma~\ref{lem:merge_order} identifies it using the set table.

\begin{lemma}[Best merge order]
\label{lem:merge_order}
With $M$ hidden, the order of $X$ that gives the
smallest $W$ can be evaluated in the following two cases.
(i) If $\mathit{Sets}[X]$ is empty, no visible prefix consists of exactly the items in \(X\), and
no node of $U_M$ can be reused. The best order follows a node of
$\mathit{Sets}[Y]$, $Y\subseteq X$, with the largest $\ell(Y)$, and then
takes the other items of $X$ in any order. This case always holds for
$M=L(X)$.
(ii) If $\mathit{Sets}[X]$ is not empty, the best order reuses all of
$X$. It is the path of a node $v\in\mathit{Sets}[X]$ with the largest
$W(U_M\cap\mathcal{T}_v)$, the weight of the nodes of $U_M$ that are
visible below $v$ with matching item sequences.
\end{lemma}

\noindent
The index thus forms one order for each move group and
evaluates it once, instead of evaluating $|X|!$ orders. Case (i) takes
at most $2^{|X|}$ lookups in the set table. Case (ii) compares only the
orders that other requests already use, which are the nodes of
$\mathit{Sets}[X]$. To compute their overlaps, the index uses the
first-item partition of Section~\ref{sec:grouping}. Case (ii) arises
only for a part of this partition, so the members of $M$ begin with the
same item $a$. If $a\notin X$, every residual order begins with $a$, so
only the nodes with a child in $\mathit{Sets}[X,a]$ have any overlap,
and any node of $\mathit{Sets}[X]$ serves when there are none. If
$a\in X$, the index searches the visible subtrees below
$\mathit{Sets}[X]$ and follows only the children that match $U_M$. The
index then restores the set table. Only the paths of the chosen order
enter the tree, so unselected orders require no additional nodes or stored paths.

\noindent\textbf{Dissolve order selection.}
Dissolve places group members one at a time, so that later members can
reuse the prefixes of earlier ones. Trying all orders of
a member would take $k!$ evaluations, but many of them cost the
same. An order reuses the items on its longest prefix that other
requests use. Every later node of the order is unused, because a request
cannot pass a node without passing its parent, so the order adds all its
other items to planned prefill.

\begin{lemma}[Equivalent orders]
\label{lem:orders}
When a request is placed, two of its orders that reuse the same items have the same $\Delta W$, and $\Delta W$ grows with the
total length of these items. The best order reuses the longest
set of items, and a request with $k$ items has at most $2^k$ such sets.
\end{lemma}

\noindent
A member thus needs only the longest set that it can
reuse. The index looks up the subsets of the member's evidence in the
set table and forms one order only.
The member must also leave $G$, so its order cannot pass
the first node of $G$. The index hides all members of $G$,
which removes the nodes of $G$ from the set table and keeps the prefixes
shared with other requests, including those of a larger group. It considers only visible prefixes that allow the request to append
all remaining evidence items without passing through the first node
of $G$. Among these prefixes, it selects the one with the largest
total token length. Only the node right above $G$
can leave no way around $G$, when the first item of $G$ is the only item
left, so the index reads at most two nodes per set. The member follows
the chosen node, continues with an item whose next node is unused and
outside $G$, and then takes its other items in their current order.
Such an item exists, since no longer reusable set was found.

Each placed path stays in the set table, so that later
members can reuse it, while $n_v$ and the layout stay unchanged. After
the last member, the index restores the set table and evaluates all the
moves together by Lemma~\ref{lem:crossing}. It accepts the dissolve only
if this $\Delta W$ is positive. If a member has no order that avoids
$G$, the dissolve is dropped. Single-request improvements search in the
same way, with only that request hidden and no group to avoid.

\begin{example}
For the layout in Figure~\ref{fig:regroup}(a), the candidate table
gives $L(\{D\})=\{r_2,r_5\}$. Hiding these requests in the set table,
the index forms the merge orders $(D,B)$ and $(D,C)$.
After the merge, it hides $r_1$ and $r_7$ to dissolve the remaining
$B$-prefix group. The visible prefixes $(A,E)$ and $(C)$ yield
the orders $(A,E,B)$ and $(C,E,B)$ in Figure~\ref{fig:regroup}(c).
The index restores the set table and evaluates the joint move.
Counts fall to zero on $(B)$, $(B,E)$, $(B,E,A)$, and $(B,E,C)$,
removing $7+1+6+4=18$ tokens. Counts rise from zero on
$(A,E,B)$, $(C,E)$, and $(C,E,B)$, adding $7+1+7=15$ tokens.
Thus $\Delta W=18-15=3$, and the accepted dissolve lowers $W$
from 53 to 50 without a full tree scan.
\end{example}


\begin{algorithm}[t]
{
\SetVline
\footnotesize
\caption{Global Prefix Index Procedures}
\label{alg:index}
\SetCommentSty{textbf}
\textbf{Procedure} $\textnormal{BuildIndex}(O)$\\
\Indp
\State{$(\mathcal{T},n,L,W,\mathit{Sets})\leftarrow\textnormal{InitializeIndex}(O)$}
\Indm
\BlankLine
\vspace{-2.51pt}
\textbf{Procedure} $\textnormal{ComputeDeltaWeight}(\mathit{change})$\\
\Indp
\State{$(V^-,V^+)\leftarrow\textnormal{ZeroCrossings}(\mathit{change},n)$}
\StateCmt{\Return $\sum_{v\in V^-}\ell(e_v)-\sum_{v\in V^+}\ell(e_v)$}{Lemma~\ref{lem:crossing}}
\Indm
\BlankLine
\vspace{-2.51pt}
\textbf{Procedure} $\textnormal{ApplyChange}(O,\mathit{change})$\\
\Indp
\State{$W\leftarrow W-\textnormal{ComputeDeltaWeight}(\mathit{change})$}
\State{$\textnormal{UpdateReferences}(\mathit{change},n,\mathit{Sets})$}
\For{$(r,o)\in\mathit{change}$}{\State{$o_r\leftarrow o$}}
\State{\Return}
\Indm
\BlankLine
\vspace{-2.51pt}
\textbf{Procedure} $\textnormal{OrderForMerge}(X,M)$\\
\Indp
\State{$\textnormal{Hide}(M,\mathit{Sets})$}
\uIf{$\mathit{Sets}[X]=\emptyset$}{
    \StateCmt{$(Y,v)\leftarrow\textnormal{LongestSubsetPrefix}(X,\mathit{Sets})$}{Max $\ell(Y)$}
    \State{$t\leftarrow\textnormal{Concat}(\textnormal{Path}(v),\textnormal{AnyOrder}(X\setminus Y))$}
}
\Else{
    \StateCmt{$v\leftarrow\textnormal{MaxOverlapNode}(X,U_M,\mathit{Sets})$}{Max overlap}
    \State{$t\leftarrow\textnormal{Path}(v)$}
}
\State{$\textnormal{Restore}(\mathit{Sets})$}
\State{\Return $t$}
\Indm
\vspace{1mm}
\textbf{Procedure} $\textnormal{PlanForDissolve}(G)$\\
\Indp

\State{$\mathit{change}\leftarrow\emptyset$}
\State{$\textnormal{Hide}(G,\mathit{Sets})$}
\For{$r\in G$}{
    \State{$v\leftarrow\textnormal{LongestFeasiblePrefix}(r,G,\mathit{Sets})$}
    \If{$v=\bot$}{
        \State{$\textnormal{Restore}(\mathit{Sets})$}
        \State{\Return $(\emptyset,0)$}
    }
    \State{$o\leftarrow\textnormal{CompleteOutsideGroup}(r,v,G)$}
    \State{$\mathit{change}\leftarrow\mathit{change}\cup\{(r,o)\}$}
    \State{$\textnormal{Insert}(o,\mathit{Sets})$}
}
\State{$\textnormal{Restore}(\mathit{Sets})$}
\State{\Return $(\mathit{change},\textnormal{ComputeDeltaWeight}(\mathit{change}))$}
\Indm
}
\end{algorithm}

\noindent\textbf{Indexed planning cost analysis.}
These searches remove the factor $k!$ from the planning
cost of Theorem~\ref{thm:cost_plain}, as Theorem~\ref{thm:cost_index}
states.

\begin{theorem}[Planning cost with the index]
\label{thm:cost_index}
For $N$ requests with at most $k$ evidence items each and at most
$I_{\max}$ rounds each of merges and dissolves, indexed planning takes
$O(I_{\max}\cdot k\cdot 3^k\cdot N+2^k\cdot N\log N)$ expected time,
excluding the greedy start and the ILP solves.
\end{theorem}

\noindent
For fixed $k$ and $I_{\max}$, regrouping is linear in $N$
after one sort of the candidates. With the $O(k^2N\log N)$ greedy start,
planning is near-linear in $N$, excluding the ILP solves for small reuse scopes.

\noindent\textbf{Algorithm description.}
Algorithm~\ref{alg:index} provides index procedures for
Algorithm~\ref{alg:grouping}. $\textnormal{BuildIndex}$ initializes their shared
index state from $O$ (lines 1--2).
$\textnormal{ZeroCrossings}$ evaluates joint count crossings without
modifying counts (lines 3--5). $\textnormal{UpdateReferences}$ commits
counts and table entries before layout updates (lines 6--11).
With moved requests hidden,
$\textnormal{LongestSubsetPrefix}$ and $\textnormal{MaxOverlapNode}$ select any
maximizer in Lemma~\ref{lem:merge_order}(i) and (ii), respectively
(lines 15 and 18). $\textnormal{AnyOrder}$ orders the remaining items
arbitrarily (line 16).
$\textnormal{PlanForDissolve}$ places hidden group members sequentially
(lines 22--34). $\textnormal{LongestFeasiblePrefix}$ maximizes reused token
length among visible prefixes admitting completion outside $G$
(line 26). $\textnormal{CompleteOutsideGroup}$ appends remaining items via an
unused node avoiding the first node of $G$, preserving their order
thereafter (line 30). An empty suffix requires no extension.
$\textnormal{Insert}$ exposes each trial path to later members (line 32).
Both exits restore pretrial state (lines 28 and 33).
Only $\textnormal{ApplyChange}$ commits moves to the layout.

\vspace{-2mm}

\subsection{Prefix-Ready Admission}
\label{sec:admission}
\vspace{-1mm}

Prefix-ready admission turns the planned
layout into execution decisions based on prefix availability. A shared
prefix avoids repeated computation only while its KV states remain
available to later requests. We submit requests in depth-first order
of the layout's prefix tree to keep members of each group close
together~\cite{zheng2024sglang}. At each node, subtrees are visited in
increasing order of their tokens per request, with requests ending at
the node placed before those continuing below it. This order improves
cache locality, but concurrent requests can still begin prefill before
their shared prefix is cached. Admission delays requests waiting on
in-flight prefixes and lets other requests run until those prefixes
are ready for reuse.

\vspace{1mm}
\noindent\textbf{Segment state tracking.}
Nested sharing requires readiness to be tracked
at several points along a prompt. Different subsets of requests may
share different lengths of its prefix, with successive parts computed
by different requests. Admission tracks each nonoverlapping segment
and its producer, allowing different parts of a prompt to become ready
at different times as their producers progress.

\begin{definition}[Segment]
\label{def:segment}
A \emph{segment} is a maximal chain of consecutive nodes on a path of
the prefix tree that the same requests pass through. Unlike prefixes,
which all start at the root and overlap, the segments of a path do not
overlap and follow one another.
\end{definition}

\noindent
A path in the prefix tree thus splits into several segments,
which are the edges of the compressed (radix) form of the tree, and all
tokens of a segment are shared by the same requests. Each segment is
computed by the first request in depth-first order
that contains it, called a producer request. Other
requests that contain the segment but share no segment after it with
any other request are its consumers. They reuse the segment and compute
only the rest of their own prompts, which no other
request shares.

\begin{example}
In Figure~\ref{fig:regroup}(c), the
$A$-prefix group runs in the order $r_4$, $r_6$, $r_1$, with
evidence orders $(A)$, $(A,E)$, and $(A,E,B)$.
Since $r_4$ ends after $A$ and $r_6$ after $E$, the path of $r_1$
splits into segments $A$, $E$, and $B$. Requests $r_4$ and $r_6$
produce the shared segments $A$ and $E$, respectively.
Once $(A,E)$ is cached, $r_1$ reuses it and computes only
its private segment $B$.
\end{example}


\begin{definition}[States of a segment]
\label{def:states}
Each segment is in one of three states:
$(i)$ ready, if its KV states are cached;
$(ii)$ in flight, if claimed by an admitted request but not yet cached; and
$(iii)$ pending, if it is neither cached nor claimed by an admitted
producer request.
\end{definition}

\begin{algorithm}[t]
	{
		\SetVline
		\footnotesize
		\caption{Prefix-ready Admission}
		\label{alg:admission}
		\SetCommentSty{textbf}
		\KwIn{a layout $O$}
		\KwOut{the requests admitted at each step}
		\State{$T\leftarrow\textnormal{SegmentTree}(O)$}
		\State{$Q\leftarrow\textnormal{DFSOrder}(T)$}
		\While(\tcc*[f]{One engine step}){$Q\ne\emptyset$}{
			\StateCmt{$\textnormal{ReleaseClaims}(T)$}{Cached or producer finished}
			\For{$s\in V(T)\setminus\{\mathit{root}\}$}{
				\lIf{$\textnormal{Cached}(s)$}{$\mathit{state}[s]\leftarrow\mathit{ready}$\;}
				\lElseIf{$\textnormal{Claimed}(s)$}{$\mathit{state}[s]\leftarrow\mathit{in\ flight}$\;}
				\lElse{$\mathit{state}[s]\leftarrow\mathit{pending}$\;}
			}
			\For{$r\in Q$}{
				\If{$\textnormal{AvailableCapacity}()=0$}{
					\State{\textbf{break}}
				}
				\If{$\forall s\in\textnormal{Segments}(r):\mathit{state}[s]\ne\mathit{in\ flight}$}{
					\State{$\textnormal{Admit}(r)$}
					\State{$Q\leftarrow\textnormal{RemoveRequest}(Q,r)$}
					\For{$s\in\textnormal{UncachedSegments}(r)$}{
						\State{$\textnormal{Claim}(s,r)$}
						\State{$\mathit{state}[s]\leftarrow\mathit{in\ flight}$}
					}
				}
			}
		}
		\State{\Return}
	}
\end{algorithm}

\begin{lemma}[Segment dependency]
\label{lem:dependency}
For a waiting request, prefill can start at segment $s$ only if
all preceding segments on its path are ready.
If a segment is pending, all its descendants are pending.
\end{lemma}

\noindent\textbf{Request holding.}
By Lemma~\ref{lem:dependency}, admission needs to check only one
segment of a waiting request, namely the first segment on its path that
is not ready. If this segment is in flight, the request would compute it
again, so the request is held until the segment is ready. If this
segment is pending, no request computes it or any segment after it,
since all segments after a pending segment are also pending. However, several
waiting requests may share this pending segment, and admission
admits only the first of them in queue order. Since requests wait in
depth-first order, this request is the producer of the segment, and it
causes no duplicate computation. For the
other requests, the segment is now in flight, so they are held.
For the layout in Figure~\ref{fig:regroup}(c), suppose $r_4$
is computing $A$ while $E$ remains pending. Both $r_6$ and $r_1$
are held because their first unready segment is $A$.
If $D$ is already cached, $r_5$ and $r_2$ can reuse it independently.

\noindent\textbf{Request admission.}
The producer of a segment enters once the segment before it is ready,
and its consumers enter when the segment itself is ready.
For the layout shown in Figure~\ref{fig:regroup}(c), once $A$ is cached, $r_6$ reuses it and computes $E$ without
waiting for $r_4$ to finish decoding. Request $r_1$ remains held
until $E$ is cached, then reuses $(A,E)$ and computes $B$.
Requests on the $C$ and $D$ branches proceed independently
of this wait.
At a branch, several segments start at the end of the same segment.
After the segment that ends at the branch is ready, the producers of the
segments after it can compute them at the same time and at different
speeds. Holding requests that wait for one segment
leaves other branches free to proceed. Instead, the producers of
pending segments enter before the held requests. On each branch, the
producer of the next segment enters as soon as the segment before it is
ready, so a slow producer on one branch does not stall the others. These readiness checks apply at every level of the tree, using
segment dependencies without tracking request groups or blocking unrelated branches of the tree.

At runtime, an admitted request claims all
segments it will compute, marking them in flight until their KV states
are cached. Admission checks the engine's prefix cache at every step,
so an evicted segment is no longer ready and may need to be computed
again.

\noindent\textbf{Algorithm description.}
Algorithm~\ref{alg:admission} summarizes prefix-ready admission on the
segments of a layout $O$. Lines 1--2 compress its prefix tree into
the segment tree $T$ and initialize the DFS queue $Q$.
Each non-root node is a segment, and the root is the empty prefix.
$\textnormal{Segments}(r)$ lists the segments on the path of $r$. Lines 3--17 follow the
engine step by step. Line 4 releases claims for cached segments or
finished producers. Lines 5--8 then refresh the segment states:
a cached segment is ready, an uncached segment with an active claim
is in flight, and every other segment is pending. The $\mathit{state}$
record reflects current cache availability and claims, so an evicted
segment is no longer ready.
Lines 9--17 scan waiting requests in depth-first order, stopping
when admission capacity is exhausted (lines 10--11). A request enters
only if none of the segments on its path is in flight (lines 12--14).
By Lemma~\ref{lem:dependency}, its path is then either fully ready or
has a pending first non-ready segment. Upon admission, the request
claims every uncached segment on its path and marks it in flight
(lines 15--17). These updates occur before the next request is checked,
so later requests in the same step see the new claims. The loop
continues until every request has entered the engine. $\textnormal{RemoveRequest}$
deletes the admitted request without reordering the queue, so each
request is admitted exactly once.

\vspace{-2mm}

\section{Experiment} 
\vspace{-1mm}
\label{sec:exp}
\begin{table*}[htbp]
\caption{End-to-end performance. Thr.: throughput (req/s, higher is better). TTFT: mean time to first token relative to KaleidoKV (lower is better). Qual.: dataset-specific answer quality (higher is better). Within each row, bold and underline mark the best and second-best values of each metric, including ties. Dashes indicate unsupported configurations or failed runs.}
\label{tab:main}
\centering
\footnotesize
\setlength{\tabcolsep}{2.6pt}
\definecolor{kvModelBlue}{HTML}{EDF4FC}
\definecolor{kvModelGreen}{HTML}{EDF7F0}
\definecolor{kvModelLavender}{HTML}{F3EFF9}
\definecolor{kvModelPeach}{HTML}{FCF2E9}
\newcommand{\kvMethodSeparator}{\hspace{1pt}{\color[gray]{0.55}\vrule width 0.35pt}\hspace{1pt}}
\begin{tabular}{ll*{15}{r}!{\kvMethodSeparator}rrr}
\toprule
 & & \multicolumn{3}{c}{ContextPilot} & \multicolumn{3}{c}{GGR} & \multicolumn{3}{c}{SOLO} & \multicolumn{3}{c}{BatchLLM} & \multicolumn{3}{c!{\kvMethodSeparator}}{SGLang} & \multicolumn{3}{c}{\textbf{KaleidoKV(Ours)}} \\
\cmidrule(lr){3-5}\cmidrule(lr){6-8}\cmidrule(lr){9-11}\cmidrule(lr){12-14}\cmidrule(lr){15-17}\cmidrule(lr){18-20}
\textbf{Model} & \textbf{Dataset} & Thr. & TTFT & Qual. & Thr. & TTFT & Qual. & Thr. & TTFT & Qual. & Thr. & TTFT & Qual. & Thr. & TTFT & Qual. & Thr. & TTFT & Qual. \\
\midrule
\rowcolor{kvModelBlue}
 & DragonBall (DB) & \underline{5.71} & \underline{1.21$\times$} & 43.1 & 4.44 & 2.01$\times$ & 43.2 & 4.59 & 1.83$\times$ & \underline{43.4} & 2.68 & 2.83$\times$ & \underline{43.4} & 3.29 & 2.37$\times$ & \textbf{43.5} & \textbf{7.06} & \textbf{1.00$\times$} & 43.3 \\
\rowcolor{kvModelBlue}
 & NarrativeQA (NQA) & \underline{3.81} & \underline{1.29$\times$} & \underline{56.3} & 3.00 & 2.05$\times$ & 56.1 & 3.15 & 1.66$\times$ & 56.1 & -- & -- & -- & 3.37 & 1.50$\times$ & 56.2 & \textbf{4.96} & \textbf{1.00$\times$} & \textbf{56.6} \\
\rowcolor{kvModelBlue}
 & SQuAD (SQ) & \underline{3.09} & \underline{1.25$\times$} & 55.5 & 2.58 & 1.94$\times$ & 55.3 & 2.60 & 1.63$\times$ & 55.5 & 2.16 & 1.97$\times$ & \textbf{56.3} & 2.39 & 1.73$\times$ & \underline{56.1} & \textbf{3.95} & \textbf{1.00$\times$} & 55.4 \\
\rowcolor{kvModelBlue}
\multirow{-4}{*}{Qwen2.5-14B-Instruct} & WikiCities (WT) & \underline{3.80} & \underline{1.86$\times$} & \underline{70.1} & 3.60 & 2.26$\times$ & 67.6 & 3.45 & 2.43$\times$ & \textbf{70.5} & 2.19 & 3.28$\times$ & 69.4 & 1.70 & 4.18$\times$ & 69.2 & \textbf{6.55} & \textbf{1.00$\times$} & 69.0 \\
\midrule
\rowcolor{kvModelGreen}
 & DragonBall & \underline{9.48} & \underline{1.26$\times$} & 43.0 & 6.63 & 2.41$\times$ & \textbf{43.3} & 7.93 & 1.74$\times$ & 42.9 & -- & -- & -- & 5.76 & 2.26$\times$ & \textbf{43.3} & \textbf{11.94} & \textbf{1.00$\times$} & \underline{43.2} \\
\rowcolor{kvModelGreen}
 & NarrativeQA & \underline{6.56} & \underline{1.32$\times$} & 56.1 & 4.61 & 2.48$\times$ & 56.2 & 5.57 & 1.62$\times$ & 56.2 & -- & -- & -- & 6.04 & 1.45$\times$ & \textbf{56.4} & \textbf{8.73} & \textbf{1.00$\times$} & \underline{56.3} \\
\rowcolor{kvModelGreen}
 & SQuAD & \underline{5.49} & \underline{1.27$\times$} & 60.2 & 4.05 & 2.37$\times$ & \underline{60.7} & 4.61 & 1.63$\times$ & 59.9 & -- & -- & -- & 4.34 & 1.69$\times$ & \textbf{60.8} & \textbf{6.98} & \textbf{1.00$\times$} & 60.4 \\
\rowcolor{kvModelGreen}
\multirow{-4}{*}{Qwen3-8B} & WikiCities & \underline{6.82} & \underline{1.70$\times$} & \underline{58.4} & 5.49 & 2.67$\times$ & 50.6 & 6.09 & 2.24$\times$ & 50.3 & -- & -- & -- & 3.14 & 3.70$\times$ & 56.6 & \textbf{11.06} & \textbf{1.00$\times$} & \textbf{59.4} \\
\midrule
\rowcolor{kvModelLavender}
 & DragonBall & \underline{10.17} & \underline{1.32$\times$} & 36.1 & 7.08 & 2.50$\times$ & 35.9 & 8.43 & 1.82$\times$ & 35.3 & 4.89 & 2.93$\times$ & \textbf{36.9} & 6.35 & 2.24$\times$ & \underline{36.8} & \textbf{12.87} & \textbf{1.00$\times$} & 35.9 \\
\rowcolor{kvModelLavender}
 & NarrativeQA & \underline{6.86} & \underline{1.38$\times$} & 51.4 & 4.79 & 2.60$\times$ & 51.5 & 5.91 & 1.66$\times$ & 51.5 & 5.67 & 1.70$\times$ & \textbf{52.2} & 6.42 & 1.47$\times$ & \underline{52.1} & \textbf{9.20} & \textbf{1.00$\times$} & 51.8 \\
\rowcolor{kvModelLavender}
 & SQuAD & \underline{5.84} & \underline{1.31$\times$} & 42.0 & 4.24 & 2.48$\times$ & \underline{42.1} & 4.89 & 1.68$\times$ & \textbf{43.1} & 4.08 & 2.03$\times$ & 41.5 & 4.61 & 1.72$\times$ & 41.7 & \textbf{7.47} & \textbf{1.00$\times$} & 41.5 \\
\rowcolor{kvModelLavender}
\multirow{-4}{*}{Llama-3-8B-Instruct} & WikiCities & \underline{7.37} & \underline{1.94$\times$} & 63.1 & 5.99 & 2.98$\times$ & \textbf{64.3} & 6.75 & 2.45$\times$ & 63.3 & 4.33 & 3.34$\times$ & 63.1 & 3.54 & 4.00$\times$ & \underline{63.4} & \textbf{12.73} & \textbf{1.00$\times$} & 63.1 \\
\midrule
\rowcolor{kvModelPeach}
 & DragonBall & \underline{10.16} & \underline{1.31$\times$} & \textbf{41.7} & 7.29 & 2.45$\times$ & \underline{41.6} & 8.45 & 1.83$\times$ & 41.0 & 4.94 & 2.90$\times$ & \underline{41.6} & 6.31 & 2.27$\times$ & \underline{41.6} & \textbf{13.02} & \textbf{1.00$\times$} & 41.3 \\
\rowcolor{kvModelPeach}
 & NarrativeQA & \underline{6.75} & \underline{1.41$\times$} & 58.1 & 4.67 & 2.69$\times$ & \underline{58.5} & 5.91 & 1.64$\times$ & 58.4 & 5.66 & 1.70$\times$ & \underline{58.5} & 6.48 & 1.45$\times$ & \textbf{58.6} & \textbf{9.21} & \textbf{1.00$\times$} & \textbf{58.6} \\
\rowcolor{kvModelPeach}
 & SQuAD & \underline{5.70} & \underline{1.31$\times$} & 58.6 & 4.33 & 2.30$\times$ & 59.0 & 4.85 & 1.63$\times$ & \textbf{59.4} & 4.07 & 1.95$\times$ & \underline{59.2} & 4.66 & 1.63$\times$ & \underline{59.2} & \textbf{7.29} & \textbf{1.00$\times$} & 58.4 \\
\rowcolor{kvModelPeach}
\multirow{-4}{*}{Llama-3.1-8B-Instruct} & WikiCities & \underline{7.16} & \underline{1.98$\times$} & 74.5 & 6.18 & 2.75$\times$ & 72.5 & 6.72 & 2.38$\times$ & 74.0 & 4.30 & 3.26$\times$ & \underline{75.3} & 3.55 & 3.84$\times$ & \textbf{75.4} & \textbf{12.52} & \textbf{1.00$\times$} & \underline{75.0} \\
\bottomrule
\end{tabular}
\end{table*}

\begin{figure*}[!t]
\centering
\captionsetup{skip=2pt,justification=raggedright,singlelinecheck=true}
\includegraphics[width=\textwidth]{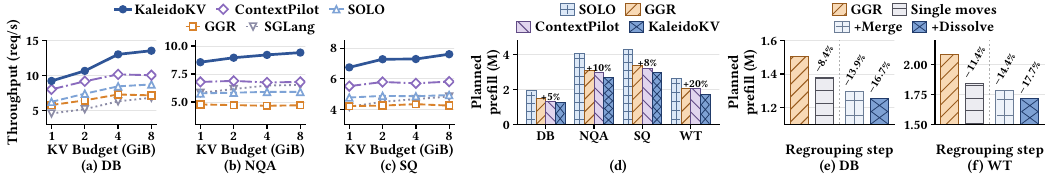}
\caption{(a-c) Throughput vs. KV cache budget. (d) Planned prefill by grouping method. (e-f) Regrouping from the GGR layout.}
\vspace{\dimexpr11pt-2mm\relax}
\label{fig:kv}\label{fig:plan}\label{fig:steps}
\end{figure*}

\begingroup
\setlength{\parfillskip}{0pt plus 0.08\linewidth}
\tolerance=1000
\emergencystretch=0.5em
\finalhyphendemerits=10000
In this section, we evaluate \sysname{} on offline RAG batches from four
question-answering datasets with four open LLMs.
Section~\ref{sec:exp:e2e} compares \sysname{} with existing
systems in end-to-end efficiency and answer quality. We then analyze
joint layout planning (Section~\ref{sec:exp:grouping}), global prefix
index (Section~\ref{sec:exp:index}), and prefix-ready admission
(Section~\ref{sec:exp:admission}). Finally,
Section~\ref{sec:exp:ablation} evaluates the contribution of each
component, and Section~\ref{sec:exp:sensitivity} studies sensitivity to the number of
evidence items. Further analyses including evidence count,
planner sensitivity, and sharing density appear in
Appendix~\supplementref{AppendixC}{C} of the online supplementary material~\cite{appendix}.
\vspace{-3mm}

\subsection{Experimental Setup}\label{sec:exp:setup}
\vspace{-1mm}

\noindent\textbf{Datasets.} We use three public QA
datasets, DragonBall (\textbf{DB})~\cite{zhu2025rageval},
NarrativeQA (\textbf{NQA})~\cite{kovcisky2018narrativeqa}, and
SQuAD (\textbf{SQ})~\cite{rajpurkar2016squad}, and one synthetic workload,
WikiCities (\textbf{WT}), with 3{,}000--3{,}461 questions each.
For public datasets, BM25 retrieves four chunks of at least
380 words per question, ordered by retrieval rank.
In public datasets, requests share chunks with only 5\%--9\% of the batch, resulting in localized sharing.
WikiCities uses comparison-style
questions~\cite{yang2018hotpotqa, ho2020constructing} over four cities,
reusing popular pages in different combinations as in entity-centric
analytics~\cite{liu2025optimizing, li2026solo}.

\noindent\textbf{Baselines.} ContextPilot~\cite{jiang2026contextpilot},
GGR~\cite{liu2025optimizing}, and SOLO~\cite{li2026solo} reorder the
evidence of each request to create shared prefixes, while
BatchLLM~\cite{zheng2026batchllm} and SGLang~\cite{zheng2024sglang} keep
the original order and share the prefixes that the prompts already
have. BatchLLM does not support Qwen3-8B and fails on
Qwen2.5-14B-Instruct with NarrativeQA.
We exclude methods such as
CacheBlend~\cite{yao2025cacheblend} that reuse KV states of non-prefix
chunks, since they may change model outputs.

\noindent\textbf{Metrics.} Throughput (Thr.) measures
requests per second over the end-to-end runtime, including planning.
TTFT is the mean time from batch submission to a request's first token,
normalized by KaleidoKV. Planned prefill counts shared prefixes once, and
actual prefill counts tokens the engine computes.
For end-to-end evaluation, both include instruction, evidence,
and question tokens. Quality (Qual.) uses ROUGE-L for DragonBall and
NarrativeQA, normalized exact match (\%) for SQuAD, and
position-balanced accuracy (\%) for WikiCities.

\noindent\textbf{Models and environment.} We use Qwen2.5-14B-Instruct,
Qwen3-8B, Llama-3-8B-Instruct, and Llama-3.1-8B-Instruct in BF16, each
served with tensor parallelism over two NVIDIA RTX A5000 GPUs. All
systems except BatchLLM run on SGLang 0.5.4 with
FlashInfer~\cite{ye2025flashinfer}, using the same requests,
prompt template, and a KV cache budget of 4\,GiB per GPU, and all
systems of a dataset and model run on the same pair of GPUs.
See Appendix~\supplementref{app:exp:setup}{D} \cite{appendix} for dataset descriptions,
baseline configurations, metric definitions, and implementation details.

\vspace{-3mm}
\subsection{Main Results}\label{sec:exp:e2e}
\vspace{-1mm}

\noindent\textbf{(RQ1) Does \sysname{} outperform existing solutions?}
To evaluate end-to-end effectiveness, we conducted
a comparison with five baselines on four datasets and four models
(Table~\ref{tab:main}). \sysname{} achieves the highest throughput and
lowest mean TTFT in all 16 settings, with 1.24--1.75$\times$ the
throughput of ContextPilot, the strongest baseline, and
1.42--3.84$\times$ that of the other systems. The gain over ContextPilot
is largest on WikiCities, at 1.62--1.75$\times$, where requests share
evidence across many groups. This pattern motivates the controlled
analyses below: joint layout planning exposes reuse, the index
limits planning cost, and admission avoids recomputing in-flight
prefixes. The mean TTFT of the baselines is 1.21--4.18$\times$ ours,
consistent with reduced batch processing time benefiting requests that
wait behind earlier ones. On the three public datasets, quality
differences from SGLang range from $-0.9$ to $+0.4$ points. The largest
decrease is on DragonBall with Llama-3-8B-Instruct. On WikiCities, the difference is at most 0.4 points
on three models, while Qwen3-8B improves by 2.8 points. Thus, the
throughput gains hold across models, although preserving the evidence
does not guarantee identical answers.

\noindent\textbf{(RQ2) Does the advantage of \sysname{} hold under
different KV cache budgets?}
To further examine sensitivity to KV cache capacity, we
conducted a budget sweep on DragonBall, NarrativeQA, and SQuAD with
Llama-3.1-8B, varying the budget from 1 to 8\,GiB per GPU
(Figure~\ref{fig:kv}(a--c)). We excluded BatchLLM because it uses a separate
engine. \sysname{} has the highest throughput at every budget. Its
throughput relative to ContextPilot increases from 1.15$\times$ to
1.35$\times$ on DragonBall, from 1.26$\times$ to 1.39$\times$ on
NarrativeQA, and from 1.22$\times$ to 1.31$\times$ on SQuAD. Larger
budgets admit more concurrent requests, increasing the opportunity to
start prefill before a shared prefix is cached. The reordering
baselines compute 7\%--19\% more prefill than planned at 1\,GiB and
12\%--38\% more at 8\,GiB. ContextPilot and GGR consequently show
limited gains or slowdowns beyond 4\,GiB. \sysname{} remains within
1.5\% of its plan, indicating that cache capacity translates into
higher throughput when admission controls duplicate prefill under
greater request concurrency.

\vspace{-3mm}

\subsection{Analysis of Joint Layout Planning}\label{sec:exp:grouping}
\vspace{-1mm}

\noindent\textbf{(RQ3) Does joint layout planning outperform baselines in planned prefill?}
To compare layout quality independently of
execution, we conducted a planned-prefill comparison on all four
datasets using the layouts evaluated in Table~\ref{tab:main}
(Figure~\ref{fig:plan}(d)). \sysname{} obtains the lowest planned prefill on every
dataset. ContextPilot plans 5\%--20\% more prefill, GGR 12\%--22\%
more, and SOLO 44\%--55\% more. The largest gap occurs on WikiCities,
where requests reuse a small pool of popular pages in different
combinations. Revising formed groups lets \sysname{} reuse this
evidence across groups.

\begin{figure}[t]
\centering
\captionsetup{skip=2pt}
\includegraphics[width=\columnwidth]{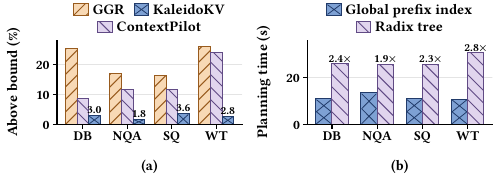}
\vspace{-4mm}

\caption{(a) Gap to the lower bound. (b) Index vs.\ radix tree.}
\vspace{-1mm}

\label{fig:bound}\label{fig:index}
\end{figure}


\newcommand{\kvMemoryPanels}{a,b,c}
\newcommand{\kvMemoryTotalPanel}{a}
\newcommand{\kvMemoryPerRequestPanel}{b}
\newcommand{\kvMemoryGrowthPanel}{c}
\begin{figure}[t]
\centering
\captionsetup{skip=2pt}
\includegraphics[width=\columnwidth]{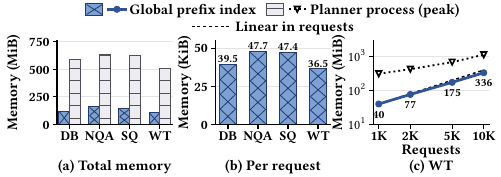}
\caption{Memory usage. (a) Total index memory and peak planner process memory at the default setting. (b) Average index memory per request. (c) Memory vs.\ batch size on WT.}
\vspace{1.5mm}

\label{fig:space}
\end{figure}

\begin{figure}[t]
\centering
\captionsetup{skip=2pt}
\includegraphics[width=\columnwidth]{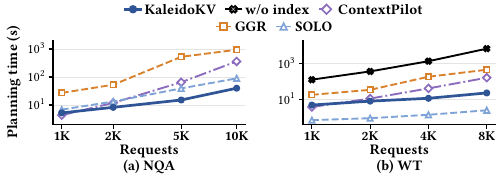}
\vspace{-4mm}

\caption{Planning time with growing batches.}
\vspace{-1mm}

\label{fig:scale}
\end{figure}

\begin{figure}[t]
\centering
\captionsetup{skip=2pt}
\includegraphics[width=\columnwidth]{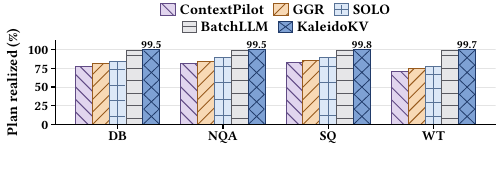}
\vspace{-10mm}
\caption{Share of the plan that each system realizes.}

\label{fig:prefill}
\end{figure}

\noindent\textbf{(RQ4) How much does each step of regrouping reduce the
planned prefill?}
To isolate the contribution of each regrouping
step, we conducted a staged comparison on DragonBall and WikiCities,
adding merge and dissolve to the same GGR layout
(Figure~\ref{fig:steps}(e,f)). A control search starts from that layout and
accepts improving single-request moves until none remains. Merge
reduces planned prefill by 13.9\% and 14.4\%, respectively, providing most of the
saving by bringing requests onto shared prefixes. Dissolve raises the
total reductions to 16.7\% and 17.7\% by removing groups whose members
can reuse longer prefixes elsewhere. In contrast, the single-request
search stops at 8.4\% and 11.4\% with similar planning time. Two
observations explain the benefit of joint layout planning. First,
creating shared prefixes provides the largest gain, but dissolving
formed groups recovers additional reuse that merge alone misses.
Second, the gap to single-request search shows the value of evaluating
moves jointly: leaving a shared prefix may be unfavorable for one
member even when moving the whole group lowers $W$.

\vspace{1mm}

\noindent\textbf{(RQ5) How close is the evidence-level cost to the optimum?}
We compare each layout's evidence-level cost $W$
with lower bounds from the ILP of Eq.~\eqref{eq:scope_ip}, solved using
HiGHS~\cite{huangfu2018parallelizing} with a 60-minute limit on four
datasets (Figure~\ref{fig:bound}(a)). HiGHS proves optimality on DragonBall and
NarrativeQA. It reaches the time limit on SQuAD and WikiCities, where
the lower bound may be below the optimum, making the reported gaps
conservative upper bounds on suboptimality. \sysname{} is within
1.8\%--3.6\% of the bound, compared with
8.6\%--23.9\% for ContextPilot and
16.3\%--26.1\% for GGR. These bounds limit the remaining reduction
in evidence-level cost on the tested batches, while repeated computation
during execution remains an avoidable runtime cost.

\vspace{2mm}

\subsection{Analysis of the Global Prefix Index}\label{sec:exp:index}

\noindent\textbf{(RQ6) Does the global prefix index plan faster than a
radix tree?}
To quantify the cost of evaluating layout changes,
we conducted an index comparison on all four datasets, replacing the
global prefix index with the radix tree used by inference
engines~\cite{zheng2024sglang} (Figure~\ref{fig:index}(b)). Both structures
return the same change in $W$, preserving planner decisions. Planning
takes 10.9--13.5~s with our index and 25.8--30.7~s with the radix tree,
a 1.9--2.8$\times$ speedup. The largest speedup occurs on WikiCities, where changes affect
the most requests. The index reduces planning time while preserving
the selected layout.

\newpage
\noindent\textbf{(RQ7) How much memory does the global prefix index use?}
{We measure the memory occupied by the global prefix
index and the peak memory usage of the entire planner process. The
index measurement includes the prefix tree and its counts, together
with the candidate, path, and set tables. 
At the default setting, the index occupies 107--161~MiB across the
four datasets, while the entire planner process peaks at
504--632~MiB (Figure~\ref{fig:space}(\kvMemoryTotalPanel)). Averaged over
the requests in each batch, the index occupies 36.5--47.7~KiB per
request (Figure~\ref{fig:space}(\kvMemoryPerRequestPanel)).
We then increase the batch size on WikiCities from 1K to 10K
requests (Figure~\ref{fig:space}(\kvMemoryGrowthPanel)). This tenfold
increase in requests increases index memory by 8.4$\times$, from
40 to 336~MiB. The average index memory per request consequently
decreases from 41.1 to 34.4~KiB. At 10K requests, the entire planner
process peaks at 1{,}140~MiB.}
\vspace{1mm}

\noindent\textbf{(RQ8) Does planning scale to larger batches?}
To test planning scalability, we conducted a
batch-size sweep on NarrativeQA and WikiCities (Figure~\ref{fig:scale}).
NarrativeQA grows from 1K requests to its full test split of 10{,}557
questions, labeled 10K, and WikiCities grows from 1K to 8K. On
NarrativeQA, planning time increases from 5.2 to 39.5~s for \sysname{}
(7.6$\times$), compared with 4.3 to 357.3~s for ContextPilot
(83$\times$), whose planner compares request pairs. On WikiCities,
\sysname{} grows by 4.7$\times$ and stays below 25~s. Recomputing $W$
without the index instead increases planning time by 54$\times$ to
6{,}972~s, consistent with Theorem~\ref{thm:cost_plain}. At the largest
batches, our planner is 9.0$\times$ faster than ContextPilot on
NarrativeQA and 7.1$\times$ faster on WikiCities. SOLO plans fastest on
WikiCities by fixing one global evidence order, but requires
1.5--1.6$\times$ our planned prefill. Therefore, planning speed must be
considered together with the layout's prefill cost and
its contribution to end-to-end runtime.

\begin{figure}[t]
\centering
\captionsetup{skip=2pt}
\includegraphics[width=.95\columnwidth]{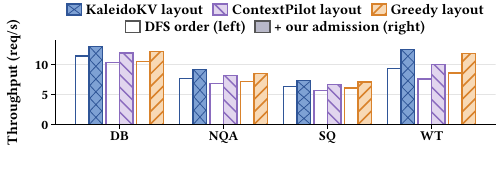}
\vspace{-1mm}
\caption{Layouts in DFS order, w/o and with our admission.}
\vspace{-1mm}

\label{fig:cross}
\end{figure}

\begin{figure}[t]
\centering
\captionsetup{skip=2pt}
\includegraphics[width=.95\columnwidth]{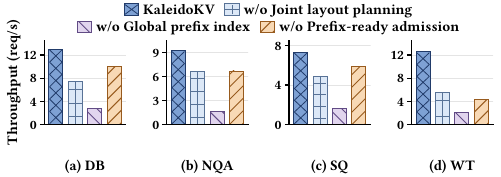}
\caption{Component ablation on the throughput.}
\label{fig:ablation}
\end{figure}

\begin{figure}[t]
\centering
\captionsetup{skip=2pt}
\includegraphics[width=0.95\columnwidth]{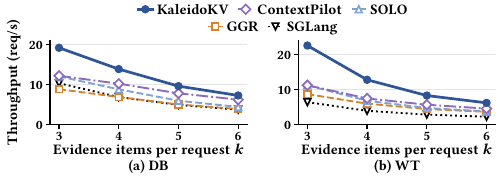}
\vspace{-1mm}

\caption{Throughput vs. evidence count.}
\vspace{-1mm}

\label{fig:kthr}
\end{figure}

\begin{figure}[t]
\centering
\captionsetup{skip=2pt}
\includegraphics[width=0.95\columnwidth]{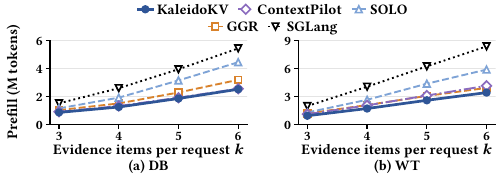}
\caption{Planned prefill vs. evidence count.}
\vspace{-1mm}
\label{fig:kplan}
\end{figure}

\vspace{-4mm}

\subsection{Analysis of Prefix-Ready Admission}\label{sec:exp:admission}
\vspace{-1mm}

\noindent\textbf{(RQ9) Does prefix-ready admission realize the planned
reuse?}
To determine whether execution realizes planned
reuse, we conducted a prefill comparison on all four datasets, measuring
planned prefill divided by actual prefill (Figure~\ref{fig:prefill}).
A value of 100\% means that actual prefill matches the plan.
\sysname{} reaches 99.5\%--99.8\%, whereas ContextPilot, GGR, and SOLO
reach 70\%--90\% because concurrent requests can repeat in-flight
prefixes. BatchLLM reaches 98\%--99\% by computing each group's shared
prefix before its members. However, its planned prefill is itself
0.7\%--17.8\% higher than the actual prefill of SGLang with DFS
submission. BatchLLM preserves retrieval order and explicitly shares
one prefix per group based on its public implementation.

\noindent\textbf{(RQ10) Does our admission also help other layouts?}
To further separate the effects of layout and
admission, we conducted a controlled comparison on all four datasets
using the layouts of \sysname{}, ContextPilot, and greedy grouping
(Figure~\ref{fig:cross}). Each layout uses the same DFS submission
order with and without prefix-ready admission. Admission increases
throughput by 1.14--1.34$\times$ for our layout,
1.15--1.32$\times$ for ContextPilot, and 1.16--1.38$\times$ for greedy
grouping. Actual prefill exceeds the plan by 17\%--40\% without
admission but remains within 0.5\% with admission. This comparison
shows that ordering requests close together is insufficient: requests
must also wait for shared prefixes that are still being computed.
With admission held fixed, our layout provides 1.09--1.25$\times$ the
throughput of ContextPilot's layout, confirming a further benefit from
joint layout planning. Together, our layout and admission provide
1.26--1.65$\times$ the throughput of ContextPilot's layout without
admission. The two controls expose complementary requirements for KV
reuse: the layout determines which computation can be shared, while
admission determines whether execution reuses it.
\vspace{-3mm}

\subsection{Ablation Study}\label{sec:exp:ablation}
\vspace{-1mm}

\noindent\textbf{(RQ11) How does each component contribute to the
end-to-end throughput?}
To assess each component's contribution, we conducted an ablation
study on four datasets with Llama-3.1-8B (Figure~\ref{fig:ablation}).
The w/o Joint Layout Planning variant preserves evidence in retrieval order.
For w/o index, we measured the CPU time to
recompute $W$ for each change and added the extra planning time to the
full system's end-to-end runtime. The variant w/o admission submits
our layout in input order and delegates admission to the engine.
Every removal lowers throughput. Without planning, planned prefill
is 1.44--2.39$\times$ that of the full system, and \sysname{} provides
1.38--2.23$\times$ the throughput of this variant. Without the index,
planning takes 881--1{,}738~s rather than 10.9--13.5~s, giving a
4.6--6.1$\times$ throughput gap. With the index, planning accounts for
less than 5\% of end-to-end time. The radix-tree control in RQ6
provides a separate comparison. Without admission, actual prefill exceeds the plan by
26.5\%--209.7\%, giving a 1.23--2.86$\times$ throughput gap. This
variant removes both DFS submission and readiness control, whereas
RQ10 isolates readiness control under fixed DFS order. The largest
losses occur on WikiCities, where extensive sharing increases both
the value of joint layout planning and the cost of redundant execution.

\vspace{-3mm}

\subsection{Parameter Sensitivity}\label{sec:exp:sensitivity}
\vspace{-1mm}

\noindent\textbf{(RQ12) How does evidence count affect performance?}
To study the effect of evidence count on
inference, we conducted a sweep on DragonBall and WikiCities with
$k=3$--6 and an 8\,GiB KV cache budget per GPU
(Figures~\ref{fig:kthr} and~\ref{fig:kplan}). WikiCities requests each
compare $k$ cities. \sysname{} has the highest throughput for every
$k$, but its ratio to the fastest baseline decreases from
1.57$\times$ to 1.17$\times$ on DragonBall and from 2.01$\times$ to
1.36$\times$ on WikiCities. Longer prompts reduce concurrency, leaving
fewer opportunities for baselines to repeat in-flight prefixes.
\sysname{} also retains the lowest planned prefill. On WikiCities,
ContextPilot plans 17.5\%--21.2\% more prefill across the sweep. On
DragonBall, its excess falls from 5.2\% to 1.3\%, while GGR and SOLO
plan 19\%--77\% more. The narrowing throughput gap reflects
both reduced duplicate computation and, on DragonBall, closer layout
quality.

\vspace{0.5mm}
To further evaluate \sysname{}, we conducted
additional experiments including the analyses of evidence count, reusable-order classes,
and regrouping rounds (Appendix~\supplementref{AppendixC}{C}). The
supplementary material~\cite{appendix} also contains omitted proofs
(Appendix~\supplementref{appendixA}{A}), the exact DP (Appendix~\supplementref{appendixDP}{B}),
and detailed experimental settings (Appendix~\supplementref{app:exp:setup}{D}).

\par
\endgroup

\vspace{-1.5mm}

\section{Conclusion} \label{sec:8}
\vspace{-1mm}

This paper studied how to reduce the prefill of offline RAG inference
by reusing the KV states of evidence shared across a batch. We
presented \sysname{}, which plans the evidence order of every request
over the whole workload and realizes the planned sharing during
execution. Joint layout planning revises groups through merges
and dissolves and keeps a change only if it lowers the prefill of the
whole workload. The global prefix index evaluates each change from the
paths of the moved requests, so that planning scales nearly linearly
with the batch. Prefix-ready admission holds only the requests whose
shared prefixes are still being computed and lets other requests run
in the meantime. Across four RAG QA datasets
and four models, \sysname{} achieves 1.24--1.75$\times$ the end-to-end
throughput of the strongest baseline and plans within 3.6\% of a
proven lower bound, without altering the retrieved evidence.

\clearpage
\bibliographystyle{ACM-Reference-Format}
\bibliography{sample}





\end{document}